\documentclass[12pt]{article}
\input epsf
\usepackage{amssymb}
\usepackage{latexsym}

\def \be{\begin{equation}}
\def \ee{\end{equation}}
\def \berr{\begin{eqnarray}}
\def \err{\end{eqnarray}}
\def \nn{\nonumber}
\def \a{\alpha}
\def \ad{\alpha^{\vee}}
\def \b{\beta}
\def \g{\gamma}
\def \d{\delta}

\def \Del{\Delta}
\def \w{\omega}
\def \L{\Lambda}
\def \Om{\Omega}

\def \eps{\varepsilon}

\def \om{\omega}
\def \l{\lambda}

\def \dl{\partial}
\def \p{\varphi}
\def\z{\zeta}
\def \tw {\tilde{\om}}

\def \A{{\cal A}}
\def \F{{\cal F}}

\def \U{{\cal U}_q}

\def \RR{{\cal R}}

\def \Fun{{\rm Fun}}

\def \({\left(}
\def \){\right)}
\def \<{\langle}
\def \>{\rangle}
\def \[{\left[}
\def \]{\right]}
\def \bb{\big|}
\def \obar{\overline}

\def\tens{\mathop{\otimes}}

\def\id{\rm id}

\newcommand \one{{\bf 1}}

\def\reps{representations }
\def\rep{representation }

\def\mf#1{{\mathbb #1}}
\def\Z{{\mf{Z}}}
\def\N{{\mf{N}}}

\def\R{{\mf{R}}}
\def\C{{\mf{C}}}

\newcommand \compl{\C}

\def\mg{\mathfrak{g}}

\def\smash{\mbox{$\,\rule{0.3pt}{1.1ex}\!\times\,$}}

\newcommand{\sect}[1]{\setcounter{equation}{0}\section{#1}}

\newtheorem{prop}{Proposition}[section]
\newtheorem{theorem}[prop]{Theorem}
\newtheorem{lemma}[prop]{Lemma}

\evensidemargin \oddsidemargin
\begin{document}

\begin{titlepage}


\begin{center}  

{\Large\bf
Quantum Anti--de Sitter space and sphere at roots of unity
 \\[4ex]}
H. \ Steinacker 
\footnote{Harold.Steinacker@physik.uni--muenchen.de}
\\[2ex] 
{\small\it 
        Sektion Physik der Ludwig--Maximilians--Universit\"at\\
        Theresienstr.\ 37, D-80333 M\"unchen  \\[1ex] }
\end{center}

\vspace{5ex}

\begin{abstract}
An algebra of functions on $q$--deformed Anti--de Sitter 
space $AdS_q^{D}$ is defined which is covariant under $U_q(so(2,D-1))$, 
for $q$ a root of unity. The star--structure is studied in detail.
The scalar fields have an intrinsic high--energy cutoff, and
arise most naturally as fields on orbifolds 
$AdS_q^{D} \times S^{D}/\Gamma$ if $D$ is odd, and 
$AdS_q^{D} \times S_{\chi}^{2D-1}/\Gamma$ if $D$ is even.
Here $\Gamma$ is a finite abelian group, and $S_{\chi}$ is a certain
``chiral sector'' of the classical sphere. Hilbert spaces of 
square integrable functions are discussed. 
Analogous results are found for the $q$--deformed sphere $S_q^D$.
\end{abstract}

\parskip 4pt plus2pt minus2pt
\vfill
\noindent
LMU-TPW 99-15

\end{titlepage}

\sect{Introduction}

The $D$--dimensional Anti--de Sitter space $AdS^D$  is a 
homogeneous space with constant negative curvature and cosmological 
constant. Its symmetry group $SO(2,D-1)$ plays the role of the
$D$--dimensional Poincar\'e group, which is recovered in the
flat limit by a contraction. It is of considerable
interest in theoretical physics for several reasons. 
For example, it can be used as a simple model for field theory on
curved spaces \cite{fronsdal}, and it arises naturally in the context
of supergravity \cite{duff}. Recently, an interesting conjecture 
relating string or M theory 
on $AdS^{D} \times W$ with (super)conformal field theories on the boundary 
has been proposed \cite{maldacena},
where $W$ is a certain sphere or a product space containing a sphere. 
Moreover, there is some evidence
that a full quantum treatment would lead to some non--classical 
version of the manifolds. This includes the appearance of a 
``stringy exclusion principle'' \cite{strominger} in the spectrum
of fields on AdS space.

In this paper, we study a non--commutative version of the AdS space,
which is covariant under the standard Drinfeld--Jimbo quantum group
$SO_q(2,D-1)$. It can be understood as a quantization of a certain
Poisson structure on the classical AdS space, where $q-1$ is a 
deformation parameter which plays the role of the Planck constant.
In principle, this is can be done for real $q$ and $q$ a phase. 
For real $q$, the qualitative features of quantum groups and spaces
are typically similar to the classical case; in particular, 
no cutoff is expected.
     
Here we consider the case where $q$ is a root of unity. It is well--known
that then quantum groups show completely new, ``non--perturbative''
features; roughly speaking, phenomena which are typical for
infinite--dimensional \reps of classical non--compact groups occur
already with finite--dimensional representations. In particular, 
it has been shown that there exist finite--dimensional unitary \reps of 
the quantum AdS groups at roots of unity \cite{nc_reps, dobrev}, 
where all the features
of the classical case are consistently combined with a cutoff.

The correct definition of quantum--AdS space for $q$ a phase
is not obvious; different versions have been proposed in the 
literature \cite{chang,ho}, which are not very satisfactory 
or incomplete. The first goal of this
paper is to clarify this situation, and to give a precise definition
in terms of operators on Hilbert spaces.
To find the proper definition, we make 2 basic assumptions: 
1) covariance under the $q$--deformed 
universal enveloping algebra $U_q(so(2,D-1))$,
and 2) allowing only finite--dimensional representations, hence 
insisting on a full regularization and avoiding ``$q$--analysis''.
It is very remarkable that this is indeed possible, 
while maintaining the correct low--energy limit.

As we will show, these assumptions lead to an algebra of functions on 
the complexified quantum sphere, which decomposes into 
different sectors corresponding to different real forms. They describe
the compact sphere $S_q^D$ and certain noncompact forms, in particular the
quantum Anti--de Sitter space $AdS_q^D$. This will provide us with scalar
fields which are unitary \reps of $U_q(so(2,D-1))$, and correspond
to the classical square--integrable scalar fields on AdS space, 
describing spin 0 elementary particles. The 
remarkable difference to the classical case is that all this 
happens within the framework of polynomial functions,
whose properties are completely different from the classical case.
Nevertheless,
the classical fields are recovered in the limit of $q$ approaching 1.
In particular, this allows to study questions of functional analysis
in the classical case with purely algebraic methods.

Moreover, it will turn out that the definition of $AdS_q^D$
implies a number of additional, unexpected features. 
They include the appearance of an additional undeformed symmetry
group $SO(D+1)$ if $D$ is odd and $Sp(D)$ if $D$ is even, which 
are in some sense spontaneously broken \cite{nc_reps}. Moreover,
it turns out that the quantum spaces are obtained most naturally as
product of the quantum AdS space (or sphere) with a classical sphere.
More precisely, one obtains  the products
$AdS_q^{2r+1} \times S^{2r+1}/\Gamma$ and 
$AdS_q^{2r} \times S_{\chi}^{4r-1}/\Gamma$, where $\Gamma = (Z_2)^r$,
and $S^{4r-1}_{\chi}$ is a certain ``chiral'' sector of $S^{4r-1}$.
The quotients of the classical spaces are actually twisted
sectors of orbifolds. It should be 
emphasized that no specific assumptions have been made here, it is simply
a consequence of the remarkable structures that appear at roots of unity.
Of course, this is quite intriguing in the context of the AdS--CFT
correspondence mentioned above, since we obtain $AdS_q^3 \times S^3,
AdS_q^5 \times S^5$ and $AdS_q^4 \times S_{\chi}^7$, which are 
precisely cases of interest there (apart from the ``chiral sector'' of $S^7$,
whose meaning is not entirely clear).
These and other physical aspects will be discussed further in Section 7.

This paper is organized as follows. In Section 2, some basic facts
about quantum groups and spaces are reviewed, including 
aspects of the \rep theory at roots of unity which
will be needed. In Section 3, we discuss in detail the meaning of reality
structures, and determine the real form of the quantum AdS group 
$\U(so(2,D-2))$. 
Section 4 is devoted to a closer analysis of the structure of 
polynomial functions on the complex quantum spaces  
at roots of unity. In Section 5, we identify 
different noncompact sectors, which leads to the definition of 
Hilbert spaces of scalar fields. Their product structure with
classical spheres is analyzed in Section 5.2. Sections 5.3 and 5.4 are 
mathematical interludes, and will allow us to write down explicitly 
the star structure
of the real quantum spaces in Theorems 5.3 and 5.4, which are some of the
main results of this work. In Section 6, we comment on further developments
towards formulating physical models, and propose an
on--shell condition which is somewhat reminiscent
of string theory. Some physical aspects are discussed in Section 7.
The Appendices include several proofs that were omitted in the 
text, as well an exposition of the vector \reps of $so(D)$ for
convenience.

Some advice to the reader: In Sections 5.3 and 5.4, the  star structure
is defined in several steps, and considerable effort is made
to give the precise mathematical definitions and to 
explain why it is the correct one. However the final result, 
Theorems 5.3 and 5.4 can be stated very briefly. 
Thus the reader who is not interested in the 
mathematical details may skip much of these sections and simply accept the
results.

\sect{The basic algebras}

We first recall the classical Anti--de Sitter space $AdS^{D-1}$,
which is a $(D-1)$ --dimensional manifold
with constant negative curvature and signature 
$(+,-,...,-)$. 
It can be embedded in a $D$--dimensional flat
space with signature $(+,+,-,...,-)$ by
\be
z_1^2+z_{D}^2-z_2^2-...-z_{D-1}^2 = R^2,  \label{ads_metric}
\ee
where $R$ will be called the "radius" of the AdS space. 
The group of isometries of this space is $SO(2,D-2)$, 
which plays the role of the $(D-1)$ --dimensional Poincar\'e group. 

This space has some rather peculiar features. Its time--like 
geodesics are finite and closed, and the time "translations"
is the $U(1)$ subgroup of rotations in the $(z_1,z_D)$ --plane.
The space--like geodesics are unbounded. There exist nice
unitary positive--energy \reps of $SO(2,D-2)$ which correspond to
elementary particles with arbitrary spin.
It is also worth recalling that $SO(2,D-2)$ is the conformal
group in $D-2$ dimensions acting on $(D-2)$ --dimensional Minkowski space,
which can be interpreted as the boundary of $AdS^{D-1}$.

To define the noncommutative version, we first review
some basic facts about
the $q$--deformed orthogonal group and Euclidean space \cite{FRT};
for a more detailed discussion see e.g. \cite{fiore,thesis}.
The algebra of functions $\Fun_q(SO(D,\C))$ on the orthogonal quantum group
is generated by matrix elements $A^i_j$ with relations 
\be
\hat R^{ik}_{mn} A^m_j A^n_l = A^i_n A^k_m \hat R^{nm}_{jl}, 
\ee
where the matrix $\hat R^{ik}_{mn}$ is explained below. 
$\Fun_q(SO(D,\C))$ is the Hopf algebra
dual to the {\em quantized universal enveloping algebra}
$U_q(so(D,\C))$, which is easier to work with in practice.
Given a root system of a simple Lie group $\mg$ with Killing metric 
$(\;,\;)$ and Cartan matrix $A_{ij}$, $\U:=U_q(\mg)$ is the Hopf algebra 
with generators $\{ X^{\pm}_i, H_i; \; i=1,...,r\}$ and 
relations \cite{jimbo,drinfeld,FRT}
\berr
\[H_i, H_j\]      &=& 0,                                   \\
\[H_i, X^{\pm}_j\] &=& \pm A_{ji} X^{\pm}_j,   \label{CR_HX}\\
\[X^+_i, X^-_j\]   &=& \d_{i,j} \frac{q^{d_i H_i} 
                          -q^{-d_i H_i}}{q^{d_i}-q^{-d_i}}
                    = \d_{i,j} [H_i]_{q_i},     
\label{UEA}
\err
plus a quantum version of the Serre relations.
Here $q$ is a complex number such that $q \neq \pm q^{-1}$,
$q_i = q^{d_i}$ where $d_i = (\a_i, \a_i)/2$ are relatively prime, and 
$[n]_{q_i} = \frac{q_i^n-q_i^{-n}}{q_i-q_i^{-1}}$ approaches $n$ 
as $q \rightarrow 1$. The comultiplication is 
\berr
\Del(H_i)          &=& H_i \tens 1 + 1 \tens H_i \nonumber \\
\Del(X^{\pm}_i) &=&  X^{\pm}_i \tens q^{-d_i H_i/2} + q^{d_i H_i/2}
\tens  X^{\pm}_i,
\label{coproduct_X}
\err
antipode and counit are
\berr
S(H_i)    &=& -H_i, \nonumber \\
S(X^+_i)  &=& -q^{d_i} X^+_i, \quad S(X^-_i)  = -q^{-d_i} X^-_i, \nonumber \\
\eps(H_i) &=& \eps(X^{\pm}_i)=0.
\err
The classical case is obtained by taking $q=1$. The consistency of this
definition can be checked explicitly.

The Cartan--Weyl involution is defined as
\be
\theta({X_i^{\pm}}) = X_i^{\mp}, \quad \theta({H_i}) = H_i,
\ee
extended as a linear anti--algebra map; in particular,
$\theta(q) = q$ for any $q \in \compl$. It is obviously consistent with the
algebra, and one can check that
\berr
(\theta\tens\theta) \Del(x) = \Del(\theta(x)), \label{theta_del}\\
S(\theta(x)) = \theta(S^{-1}(x)).
\err
Borel subalgebras $\U^{\pm, 0}$ can be defined in the obvious way. 
This defines a quasitriangular Hopf algebra,
which means that there exists a special element 
${\cal R} \in \U \tens \U$ which satisfies 
\be
\Del'(x) = \RR \Del(x) \RR^{-1}                 \label{qtr}
\ee
for any $x \in \U$, and other properties which will not be
used explicitly. Here $\Del'(x) = \tau \circ \Del(x)$ is the 
flipped coproduct. There are explicit formulas for $\RR$, of the form
\cite{tolstoi,kirill_resh}
\be
\RR=q^{\sum  \a_{ij}^{-1} d_i H_i\tens d_j H_j}
  \( 1\tens 1 + \sum \U^{+ res} \tens \U^{- res}\)
\label{R_formula}
\ee 
where $\a_{ij} = (\a_i,\a_j)$.
In this paper, we consider $\mg = so(2r+1) = B_r$ and 
$\mg = so(2r) = D_r$.

As was shown in \cite{drinfeld}, the following remarkable element  
\be
v = (S\RR_2) \RR_1 q^{-2\tilde{\rho}}
\label{v}
\ee
is in the center of $\U$, and will be called Drinfeld--Casimir.
Here $\tilde{\rho}$ is dual to the Weyl vector 
$\rho = \frac 12 \sum_{\a>0} \a$. It satisfies
\berr
\Del(v) &=& \RR^{-1} \RR_{21}^{-1} v\tens v, \label{v_coprod}  \\
v^{-1} &=& q^{2\tilde{\rho}} \RR_2 S^2(\RR_1), \label{v_inv} 
\err 
where $\RR_{12} = \RR$ and $\RR_{21} = \tau \circ \RR$.

Consider the matrix $R^{ij}_{kl} = \pi^i_k \tens \pi^j_l(\cal R)$
where $\pi^i_k$ is the $D$--dimensional 
(``vector'') \rep of $\U$, and let 
$q_S = q$ for $so(2r)$, and\footnote{ This is 
necessary to be compatible with $\U$, as will
be checked below. In \cite{FRT}, the conventions
are such that $q_S$ is replaced by $q$.}
$q_S=q^2$ for $so(2r+1)$.
Then $\hat R^{ij}_{kl} = R^{ji}_{kl}$ decomposes as
$\hat R^{ij}_{kl} = (q_S P^+ - q_S^{-1} P^- +q_S^{1-D} P^0)^{ij}_{kl}$,
where $P^+$, $P^-$ and $P^0$ are the invariant projectors on the traceless
symmetric, the antisymmetric, and the singlet component in the 
tensor product of 2 vector representations, respectively.
The invariant tensor $g^{ij}$ is given by
$(P^0)^{ij}_{kl} =\frac {q_S^2-1}{(q_S^D-1)(q_S^{2-D}+1)} g^{ij} g_{kl}$
where $g_{ik}g^{kj} = \delta_i^j$, and can be normalized 
such that \cite{FRT}
\be
g^{ij} = g_{ij} = \d_{i,j'} q^{-\rho_i}            \label{metric}
\ee
where $j'=D+1-j$, and $\rho_i$ is given in Appendix A. 

There is a canonical way to generalize the 
classical algebra of coordinate functions on $\C^D$.
The algebra $\C_q^{D}$ defining the
{\em complex quantum Euclidean space} \cite{FRT} is
generated by $x_i$, with commutation relations
\be
(P^-)^{ij}_{kl} x_i x_j =0
\label{euclid_alg}
\ee
which are invariant under $\U$. Its center is generated by 1 and 
\be
x^2 := g^{ij} x_i x_j.
\ee
The {\em complex quantum sphere} $S_{q,\C}^{D-1}$ is generated by 
$t_i$ which satisfy the same relations (\ref{euclid_alg}) as $x_i$, 
and in addition \cite{FRT}
\be
t^2 = g^{ij} t_i t_j = 1.
\label{t_2}
\ee
Explicitly, the commutation relations for $x_i$ (and $t_i$) are \cite{FRT}
\berr
x_i x_j       &=& q_S x_j x_i, \quad \quad  i \neq j',\; i < j   \nn\\ \
[x_{i'}, x_i] &=& (q_S^2-1)\sum_{j=1}^{i'-1}q^{\rho_{i'} - \rho_j} 
               x_j x_{j'} - \frac{q_S^2-1}{1+q_S^{D-2}}\; q^{\rho_{i'}} x^2,
               \quad i < i'
\label{euclid_CR}
\err
Both algebras are covariant under the
right coaction $x_i \rightarrow x_j\tens A^j_i$ of $\Fun_q(SO(D,\C))$, 
which is equivalent to a left action 
\be
x_i \rightarrow u\cdot x_i = x_j \pi^j_i(u)              \label{action}
\ee
for $u\in U_q(so(D,\C))$. We will usually work with the latter,
which is more familiar from the classical Lie algebras; then
$\pi^i_j(u u') = \pi^i_k(u) \pi^k_j(u')$.
We can use this for a quick check of the first relation in (\ref{euclid_CR}):
\be
X_1^+ \cdot (x_1 x_2 - q_S x_2 x_1) = 
  q_S^{1/2} x_1 x_1 - q_S q_S^{-1/2} x_1 x_1 = 0
\ee
using (\ref{coproduct_X}) and Appendix A,
as it must be. The other relations can be checked similarly.

The algebra of functions $AdS_q^{D-1}$ defining the quantum 
Anti--de Sitter space  
will be defined as a real form of this
complex quantum sphere, with a (co)action of
the quantum Anti-de Sitter group. 
Therefore as an algebra, it is again defined by 
$(P^-)^{ij}_{kl} t_i t_j = 0$, $t^2 = 1$.
One could introduce a physical scale by setting
\be
y_i := t_i R                    \label{ads_y}
\ee
for a constant $R >0$,
so that $y^2 = R^2$. We will simply use the units defined by 
$R=1$; physically speaking, the scale will be set by the ``radius''
of AdS space.

So far, all these spaces are complex. The crucial issue is to find the 
correct definition of the corresponding {\em real} quantum spaces.
This is not obvious especially if $q$ is 
a phase, and in fact different possibilities have been 
proposed in the literature \cite{chang,ho}. This will be discussed in detail in later
sections. 

\subsection{Roots of Unity and Representations}
\label{subsec:reps}

Since we are primarily interested in the case where $q$ is a root of unity, 
we will consider a more powerful version of the above, the
so--called ``restricted specialization'' $\U^{res}:= U_q^{res}(so(D,\C))$ 
\cite{lusztig} with generators 
$X_i^{\pm(k)} = \frac{(X_i^{\pm})^k}{[k]_{q_i}!}$ 
for $k \in \N$ as well as $H_i$. For generic $q$,
i.e. $q$ not a root of unity, this is the same as before.
However if $q$ is a root of unity, the situation is very different.
With hindsight, we restrict ourselves to roots of unity of the form
\be
q=e^{i \pi /M},                                     \label{q_root}
\ee
where $M$ is even if $D$ is odd;
the reason for this will become clear later. Then $k=M$ is
the smallest positive integer such that $[k]_q =0$, since 
$q^{M} = -1$. We also define $M_i = M/d_i$, and $M_{\a} = M/d_{\a}$ 
for any root $\a$. Then $M_i$ is the smallest integer such that
\be
[M_i]_{q_i} =0,
\label{M_i}
\ee
since $q_i= e^{i \pi d_i/M}$.

The important point is that the additional generators $X_i^{\pm (M_i)}$ 
in $\U^{res}$ are nevertheless well--defined with
a well--defined coproduct, and therefore act on
tensor products of representations. 
In particular,
$(X_i^{\pm})^{M_i}=0$ in $\U^{res}$. Therefore
$\U^{res}$ contains a remarkable sub--Hopf algebra
$\U^{fin}$ (the ``small quantum group'') generated by
$X_i^{\pm}$ and  $H_i$. Here we have 
slightly changed the standard convention and 
included the $H_i$ as well, slightly abusing the name ``finite''. 
This is more appropriate for our purpose.


We will only consider finite--dimensional \reps in this paper. 
The Cartan generators can then be diagonalized, with eigenvalues
\be
<H_i,\l> = \frac {(\a_i,\l)}{d_i} = (\ad_i,\l),
\ee
on a weight $\l$, where as usual $\ad = \frac{2\a}{(\a,\a)}$ is the coroot of $\a$.
The fundamental weights $\L_i$ satisfy $(\L_i, \ad_j) =  \d_{i,j}$,
therefore
\be
<H_i,\L_j> = \d_{ij},
\ee
and span the lattice of integral weights. 
The irreducible highest--weight \reps with highest weight $\l$
will be denoted by $L^{res}(\l)$.

The vector \rep $V_{D}$ of $\U(so(D,\C))$ is the
\rep $L^{res}(\l_1)$ with basis $x_i$ (or $t_i$) for $i=1,...,D$. 
Their weights $\l_i$ are given explicitly in Appendix A.
If $D>4$, then the highest weight $\l_1$ is equal to $\L_1$. We also define
\be
d_S = (\l_1,\l_1), \quad M_S = M/d_S, 
\ee  
so that $q_S = q^{d_S}$ which was used above.

It is well--known \cite{rosso} that for generic 
$q$, the representation theory is essentially the same 
as in the classical case. In particular, all finite--dimensional 
\reps (=modules) of $\U^{res}$ are direct sums of some
$L^{res}(\l)$. Their character 
\be
\chi (L^{res}(\l)) = e^{\l}\sum_{\eta >0}\dim{L^{res}(\l)_{\eta}}e^{-\eta}
  =:\chi(\l)
\label{char}
\ee
is given by Weyls formula. Here $L^{res}(\l)_{\eta}$ is the weight space
of $L^{res}(\l)$ with weight $\l-\eta$.
The irreducible highest weight \reps of $\U^{fin}$ will be denoted by 
$L^{fin}(\l)$.

The value of the Drinfeld--Casimir $v$ (\ref{v}) on $L^{res}(\l)$ 
(and on any highest--weight module with highest weight $\l$)
was first determined in \cite{resh_1}:
\be
v \cdot w = q^{-c_{\l}} w \quad \mbox{for} \;\; w \in L^{res}(\l),
\label{v_h_w}
\ee
where $c_{\l}=(\l,\l+2\rho)$ is the value of the {\em classical} 
quadratic Casimir on $L(\l)$. In particular for highest weights of the 
form $\l = k \l_1$, the classical Casimir for $so(2r+1)$ is 
\be
c_{k \l_1} = 2k^2 + 2k(D-2),
\ee
and for $so(2r)$ it is
\be
c_{k \l_1} = k^2 + k(D-2).
\ee

Finally, we quote a few important facts about irreducible
\reps at roots of unity. The first one \cite{anderson,CH_P} 
states that the structure of $L^{res}$ with ``small'' highest weight $\l$ 
is the same as classically:

\begin{theorem}
Assume that  $\l$ is a dominant integral weight with
$(\l + \rho, \a) \leq M$ for all positive roots $\a$.
Then the highest weight \rep $L^{res}(\l)$ has the same character 
$\chi$ as in the classical case, given by Weyl's character formula.
\label{char_thm}
\end{theorem}
This follows from the strong linkage principle, 
which was first shown in \cite{anderson}; 
for a more elementary approach, see \cite{ads_paper}.
Moreover, $L^{fin}(\l) = L^{res}(\l)$
for these weights $\l$, since the $X_{i}^{\pm (M_{i})}$ act trivially.

For general $\l$, the structure of $L^{res}(\l)$ is difficult to analyze.
However for the ``special weights''
\be
\l_z = \sum M_i z_i \L_i   \quad \mbox{for } \; z_i \in \Z
\label{l_z}
\ee
it can be understood easily, and this will be the key for 
much of the following. 
The relation (\ref{UEA}) together with (\ref{M_i})
implies that for any highest weight module $\U^{res} \cdot w_{\l_z}$ 
with highest weight $\l_z$,
\be
\p_i:= X_i^- \cdot w_{\l_z}
\ee
is a highest weight vector (possibly zero) for any $i$, i.e. 
$X_i^+ \cdot \p_j=0$ 
for all $i,j$. Because $L^{res}(\l_z)$ is irreducible by definition, 
it follows that
\be
X_i^- \cdot w_{\l_z}=0 \;\; \mbox{for all} \; i
\label{X_i_0_L}
\ee
in  $L^{res}(\l_z)$.
In particular, the irreducible \reps $L^{fin}(\l_z)$
of $\U^{fin}(so(D))$ are one--dimensional. 
However the ``large'' generators $X_{i}^{\pm (M_{i})} \in \U^{res}$ 
do act nontrivially on $L^{res}(\l_z)$, as we will see next.

Using the commutation relations, any element of 
$\U^{-res}(so(D))$ can be written as a sum of terms of the form 
$X_{i_1}^{-(M_{i_1})} ...\; X_{i_k}^{-(M_{i_k})}\; \U^{-fin}$. 
It follows that all weights of $L^{res}(\l_z)$ have the form 
$\l_{z'} = \l_z - \sum_i n_i M_i \a_i$ with $n_i \in \N$.
In other words, $L^{res}(\l_{z})$ is a direct sum of 
one--dimensional \reps $L^{fin}(\l_{z'})$ of $\U^{fin}$, since
all $\l_{z'}$ are special points. In fact, the weights $\l_z$ have the 
structure of a weight lattice 
with ``fundamental weights'' $M_i \L_i$. This turns out to be 
the rescaled lattice of a dual Lie algebra $\tilde \mg$,
with Cartan matrix $\tilde A_{ij} = A_{ji}$  provided (\ref{q_root}) holds 
\cite{nc_reps}. In the present case, $\tilde \mg = so(D)$ if $D$ is even, 
and $\tilde \mg = sp(D-1)$ if $D$ is odd.
In fact, $\U^{res}$ ``contains'' a corresponding classical Lie algebra
as a quotient.
This is the essence of a remarkable result of Lusztig \cite{lusztig_book},
and can be made explicit as follows (\cite{nc_reps}, Theorem 4.2):

Let $a_i \in \{0,1\}$ such that 
$a_i + a_j = 1$ if $A_{ij} \neq 0$ and $i\neq j$;
this is always possible. Define $\tilde K_i = q^{d_i M_i H_i}$, and
\berr
\tilde X_i^+ &=& X_i^{+(M_i)} \tilde K_i^{a_i},\nn\\
\tilde X_i^- &=& X_i^{-(M_i)} \tilde K_i^{1-a_i}q_i^{M_i^2},\nn\\
\tilde H_i   &=& [\tilde X_i^+,\tilde X_i^-]
\label{class_generators}
\err
Then one can show the following:
\begin{theorem} 
For all special weights $\l_{z}$, 
$L^{res}(\l_z)$ is an irreducible highest--weight \rep of the classical
universal enveloping algebra $U(\tilde \mg)$, with generators
$\tilde X_i^{\pm}$ and $\tilde H_i$. 
If $v_{z'} \in L^{res}(\l_{z})$ has weight $\sum_j z'_j M_j \L_j$,
then $\tilde H_i \cdot v_{z'} = z'_i v_{z'}$.
\label{classical_algebra}
\end{theorem}
In particular if $\l_z$ is a dominant weight, then the character of 
$L^{res}(\l_z)$ is invariant under the Weyl group, and can be obtained from
Weyl's formula by rescaling the weights accordingly.

Using this, the structure of $L^{res}(\l)$ with ``large'' dominant integral
$\l$ can be described as follows \cite{CH_P}:
\begin{theorem}
Let  $\l_z$ as in (\ref{l_z}) and  $\l_0$ be an integral weights with 
$0 \leq (\l_0,\ad_i) < M_i$ for all $i$. Then
\be
L^{res}(\l_0 + \l_z) = L^{res}(\l_0) \tens L^{res}(\l_{z}).
\label{L_res_l}
\ee
\label{L_res_theorem}
\end{theorem}
This is not hard to prove, see \cite{CH_P} or \cite{nc_reps}.
Moreover, the generators (\ref{class_generators})
essentially act on the second factor in (\ref{L_res_l}) 
and $\U^{fin}$ on the first, but with a certain ``twisting'' \cite{nc_reps}.

\sect{Reality structures and symmetry algebra} 

Since the proper choice of the reality structure is crucial in the
following, we first discuss the relation
of the real structures on the spaces with their symmetry algebras.

The algebra of functions on both classical and quantum $D$--dimensional 
complex Euclidean space is generated by
coordinate functions $x_i$, which transform in the 
vector \rep $V_D$ of $U_q(so(D,\C))$. The 
tensor product of 2 such \reps contains a unique trivial representation;
in other words, there is an invariant bilinear form
$<\;,\;>: V_D \times V_D \rightarrow \C$. 
In Hopf--algebra language, invariance means
$<f,g> = <u_{(1)} \cdot f,u_{(2)}\cdot g>$ for $f,g \in V_D$,
where $\Del(u) = u\tens 1 + 1\tens u = u_{(1)}\tens u_{(2)}$
denotes the coproduct of $u\in U(so(D,\C))$ or $\U$.
This extends immediately to polynomial functions.

In the classical case, the algebra of complex functions on {\em real}
Euclidean space (or any real manifold)  is equipped 
with a $*$--structure, i.e. an antilinear map whose square is the
identity, defined by the complex conjugation.
The above invariant bilinear form then induces
a {\em hermitian} inner product by 
\be
(f,g) = <f^*,g>,
\label{inner_prod}
\ee
which satisfies $(f,g)^* = (g,f)$. The signature $(p,D-p)$ 
of this inner product 
is the signature of the (pseudo)Euclidean real space, and it is
positive definite only in the Euclidean case. But in any case, it
defines a $\ast$--structure on the algebra
$U(so(D,\C))$,  i.e. an antilinear anti--algebra map, by 
\be
(f,u\cdot g) = (u^*\cdot f,g)
\label{invariant}
\ee
for all $u \in U(so(D,\C))$. This defines the real form $U(so(p,D-p))$,
and one can then study its unitary representations, which are
infinite--dimensional except in the euclidean case.

In the $q$--deformed case, we use this connection between the real form
of a function space and its symmetry algebra in the other direction:
there will be a clear choice of the real form of $\U$,
which then determines the real form of the function algebras.
A {\em real form} or {\em $\ast$ --structure} of $\U$  
is an antilinear anti--algebra map $\ast$ on $\U$
whose square is the identity.

An {\em invariant inner product} on a \rep of $\U$ is a hermitian 
inner product  which satisfies (\ref{invariant}) for all $u \in \U$;
in other words, the star on  $\U$ is implemented 
by the adjoint, which is well--defined for nondegenerate inner products.
This is particularly intuitive for 
elements of $\U$ of the form $g=exp(itu)$ with $u^* =u$, since then
$g^* = g^{-1}$.

A \rep of $\U$ is {\em unitary} with respect to a real form of $\U$
if it has a positive definite invariant inner product. Then
\be
(\pi(u)_i^j)^* = \pi_j^i(u^*)
\label{unitary_basis}
\ee
for any $u\in \U$, in an orthonormal basis.

Our guiding principle to find the appropriate 
$q$--deformed real spaces is that the star structure on $\U$ should admit a
sufficiently large class of unitary \reps of the quantum AdS group
in order to describe elementary particles, in the spirit of Wigner. The 
AdS group is particularly well suited for such an approach, because it has
unitary \reps for any half--integer spin corresponding to  
massive as well as massless particles with positive energy, 
in any dimension. 
In fact, one can choose the Cartan subalgebra such that the energy 
is one of its generators, and the unitary \reps are then lowest--weight 
\reps with positive energy and discrete spectrum. 
The unitary \reps of the Poincar\'e group are recovered in the flat limit.
We want to maintain these features in the $q$--deformed case.
This will uniquely select the real structure.

\subsection{Star structures on $\U$}
\label{subsec:star_U}

There are essentially 2
types of star structures\footnote{There are additional ones if the
Dynkin diagram has automophisms, such as for $so(2r)$; they however do not 
correspond to Anti--de Sitter groups, and do not admit finite--dimensional 
unitary representations.} on $\U$  \cite{nc_reps}, 
the first of the form
\be
{X_i^{\pm}}^* = s_i X_i^{\mp}, \quad H_i^* = H_i,
\label{first}
\ee
and the second of the form
\be
{X_i^{\pm}}^* = s_i X_i^{\pm}, \quad H_i^* = -H_i,
\label{second}
\ee
with $s_i = \pm 1$. They define consistent star algebras 
for both real $q$ and $q$ a root of unity.
The compatibility conditions with the coproduct are different for 
real $q$ and $q$ a phase, which will show up in different reality structures
of the associated quantum spaces.

It is known that there exist finite--dimensional unitary \reps 
of the first type (\ref{first}) if $q$ is a root of unity 
\cite{ads_paper,nc_reps}, which have the desired properties 
including a high--energy cutoff in the $AdS$ case.
For $q \in\R$, unitary \reps of noncompact forms also exist, 
but they have no cutoff.
There are also certain unitary \reps of the second type for $q$ a phase,
e.g. for $U_q(so(2,1))$ \cite{schmuedgen}, but they are not 
highest--weight representations; 
in particular, the Cartan subalgebra cannot be diagonalized, and 
the energy in the AdS case is not positive definite 
(notice that the Cartan subalgebra 
is distinguished for $q\neq 1$, unlike in the classical case).
Finite--dimensional unitary representations of the second type
cannot exist, since then $H_i$ could be diagonalized with purely
imaginary spectrum, which is in contradiction with the commutation relation 
(\ref{CR_HX});
Therefore we only consider star structures
on $\U$ of the first type (\ref{first}) from now on, with
$q$ a root of unity.

Consider the vector \rep $V_{D}$ of $\U$, with basis $x_i$ 
and weights $\l_i$ for $i=1,...,D$. 
$V_D$ is unitary with respect to the compact form
\be
{X_i^{\pm}}^* =  X_i^{\mp}, \quad H_i^* = H_i,
\label{euclid_involution}
\ee
which defines $U_q(so(D))$ for both real $q$ and $q$ a phase.
In general, there is a unique invariant inner product on highest weight modules
satisfying (\ref{invariant}) for star structures of 
the type (\ref{first}). The weight vectors $x_i$ are
then orthogonal for different weights, 
and they can defined to be orthonormal, i.e. 
\be
(x_i,x_j) = \d_{ij}.
\label{euclid_inner}
\ee
This is the standard convention in the literature. 

Now it is easy to find the definition of $U_q(so(2,D-2))$.
Let $E$ be the element of the 
Cartan subalgebra which is dual to $\l_1/d_S$, so that 
$<E,\l> = (\l_1,\l)/d_S$. Then the eigenvalues of $E$ on $V_{D}$ 
are $E_i:=(\l_1,\l_i)/d_S = (1,0,...,0,-1)$; $E$ 
will turn out to be the energy.
We claim that the star structure defining $U_q(so(2,D-2))$ is 
\be 
{X_i^{\pm}}^*  =(-1)^E \theta(X_i^{\pm}) (-1)^E  = s_i X_i^{\mp}, 
    \quad H_i^* = H_i \qquad \mbox{with} \;\; s_i = (-1)^{<E,\a_i>},
\label{ads_involution}
\ee
where $s_i = (-1,1...,1)$ for $D \neq 4$, and $s_i = (-1,-1)$ for $D=4$.
This is a star algebra of the first type (\ref{first}) which
can be considered for both $q \in \R$ and $q$ a root of unity, 
and there exist unitary \reps in both cases. 
The maximal compact subalgebra is  $U_q(so(D-2)) \times U_q(so(2))$
where $U_q(so(D-2))$ generated by 
$X_2^{\pm}, ..., X_r^{\pm}, H_2, ..., H_r$ (for $D\neq 4$), 
and $U_q(so(2))$ by $E$; these subalgebras commute.
The unique (up to normalization) corresponding invariant inner product 
on $V_{D}$ which satisfies (\ref{invariant}) is
\be
(x_i,x_j) = (-1)^{E_i} \d_{ij}
\label{ads_inner}
\ee
in the above basis, and has the correct signature for the $AdS$ case. 
We will also find the desired unitary \reps of $U_q^{fin}(so(2,D-2))$,
provided (\ref{q_root}) holds.
Therefore we define the ``physical'' quantum Anti--de Sitter group at 
roots of unity to be the real form (\ref{ads_involution}) of 
$U_q^{fin}(so(2,D-2))$.

\subsection{Quantum Euclidean and Anti--de Sitter space for real $q$}

For real $q$, there is a standard star structure on the 
algebra $\C_q^{D}$. The algebra 
$\R_q^{D}$ on the {\em real quantum Euclidean space} is defined by \cite{FRT}
\be
x_i^* = x_j g^{ji}.
\label{euclid_star}
\ee
This is a consistent star algebra, which indeed leads to 
the invariant inner product $(x_i,x_j) = \d_{ij}$,
and by (\ref{inner_prod}) corresponds to the real form
(\ref{euclid_involution}) on $\U$. It is also consistent 
with the constraint $x_i x_j g^{ij} =1$, thereby defining the 
{\em quantum Euclidean sphere} $S_q^{D-1}$ for real $q$.

The algebra $AdS_q^{D-1}$ on {\em quantum Anti--de Sitter space}
for real $q$ is similarly defined by 
\be
t_i^* = - (-1)^{E_i} t_j g^{ji},
\label{ads_real}
\ee
together with 
\be
t^2 = t_i t_j g^{ij} =1.
\ee
This is consistent with (\ref{euclid_alg}) because $E$ is in the 
Cartan subalgebra, and has the correct classical limit. 
By (\ref{inner_prod}), this leads to the invariant inner product 
$(t_i,t_j) = (-1)^{E_i} \d_{ij}$. 

If $q$ is a phase, (\ref{euclid_star}) 
does not extend as a star on the algebra $\C_q^{D}$.
In this case, star structures different from  
(\ref{euclid_star}) have been proposed, of the type $x_i^* = \pm x_i$
\cite{FRT,chang}. While $\C_q^D$ then becomes a star algebra,
they lead to star structures on $\U$ of the second type (\ref{second}),
which we have discarded above. 
However, we will see below that there is a star on the semidirect product
algebra $\U \smash \C_q^{D}$ which extends (\ref{ads_involution}),
and agrees with (\ref{euclid_star}) or (\ref{ads_real}) 
in the classical limit.

\sect{Scalar fields at roots of unity}

In order to construct scalar fields on classical AdS space which are unitary
representations, one has to consider square--integrable functions;
they are of course not polynomials. The situation for $q \in \R$ is
completely analogous. At roots of unity however, the 
structure of polynomials  changes 
completely, and we will see  that the analogues of the classical 
scalar fields can in fact be written as polynomials in the $t_i$.

To see this, we have to study $\C_q^D$ in more detail.
Consider the set of homogeneous polynomials $M_q^k \subset \C_q^D$ 
with degree $k$ in the $x_i$, 
which forms a submodule of the $k$--fold tensor product \rep $V_D^{\tens k}$ 
of $\U^{res}$. $M_q^k$  is not irreducible, because the metric
projector may be nontrivial. Clearly $x_1 ... x_1 = (x_1)^k \in M_q^k$ 
is a highest weight vector with weight $k \l_1$. It generates the
highest weight module
\be
\F(k):= \U^{res} \cdot (x_1)^{k}  \subset M_q^k.      \label{F_k}
\ee
If $q$ is generic, then $\F(k)$ is a irreducible \rep
with highest weight $k\l_1$ 
corresponding to a totally ($q$--)symmetric traceless tensor, and 
$M_q^k = \F(k) \oplus x^2 \F(k-2) \oplus ...$ as classically.
If $q$ is a root of unity, then $M_q^k$ is not completely reducible any more,
which is a typical phenomenon at roots of unity.  
The full structure of $M_q^k$ is quite complicated, 
and will be discussed elsewhere. Here we only consider those modes
which will be needed for the Hilbert space of scalar fields on $AdS_q$. 

First, we identify the polynomials $\F(k)$ in $\C_q^D$ which have essentially
the same structure as classically, which means the character of
$L^{res}(k \l_1)$ is the same as classically.
Using (\ref{q_root}), Theorem \ref{char_thm} implies that 
$\chi(L^{res}(k \l_1)) = \chi(k \l_1)$ for $k \leq M_S-(D-3)$ (assuming
$D \geq 4$; indeed, any positive root $\a$ can then be written as 
$\a = \sum n_i \a_i$
with $n_i \leq a_i$, where $a_i$ are the Coxeter labels. Hence
$(k \l_1 + \rho, \a) \leq (k + D-3)d_1$, which is $\leq M = d_1 M_1$ 
provided $k \leq M_1-(D-3)$. For $D=3$, the bound is $k \leq M_S - 1/2$).

However, this bound can be improved using the strong linkage principle, 
see e.g. \cite{ads_paper}. We can assume that $k < M_S$; then it
implies that the character of $L^{res}(k\l_1) = L^{fin}(k\l_1)$ is
the same as classically, unless it contains a dominant integral weight $\mu$
which is in the orbit of $k \l_1$ under the Weyl group acting with  
center $M_S \l_1 -\rho$ (since $M_s \l_1$ is a special weight). 
Since $(\frac{\l_1}{d_S}, \rho) = <E,\rho> = (D-2)/2$
by Appendix A, this is only possible for $k > k_S$ where 
\be
k_S = M_S - (D-2)/2.
\label{k_S} 
\ee 
Hence 
\be
\chi(L^{res}(k \l_1)) = \chi(k \l_1) \quad \mbox{for}\;\; k \leq k_S.
\ee
In fact, it turns out that the smallest $k$ where $\chi(L^{res}(k \l_1))$
is non--classical for even $D > 4$ is $k=k_S+1$, where
\be
\chi(L^{res}((k_S+1) \l_1)) = \chi((k_S+1) \l_1) - \chi((k_S-1) \l_1).
\ee
(For odd $D$, there is a similar phenomenon).
This is related to the scalar singleton field on the AdS space, as we will
see in the next section. One can check that e.g. the Casimir $v$ 
(\ref{v}) indeed becomes degenerate for these weights.

To summarize, the structure of $M_q^k$ is the 
same as classically for small $k$:

\begin{theorem}
For $n_0 \leq k_S$ and roots of unity of the form (\ref{q_root}),
\be
\F(n_0) \cong L^{res}(n_0 \l_1) = L^{fin}(n_0 \l_1)
\ee
has the same character $\chi(n_0\l_1)$ as classically. Moreover, 
$M_q^{n_0}$ is the direct sum
\be
M_q^{n_0}  = \bigoplus_{0 \leq k \leq n_0/2} \;\; (x^2)^k \F(n_0-2k).
\label{M_k_decomp}
\ee
\label{scalar_thm}
\end{theorem}
The proof is completed in Appendix C.   $\Box$


Next  consider $\F(k M_S)$, which is of central importance to  us.
Its highest weight $k M_S \l_1$ is a special weight (\ref{l_z}).
Using (\ref{coproduct_X}) and the commutation relations (\ref{euclid_CR}), 
one has 
\berr
X_1^- \cdot (x_1)^{n} &=& q_S^{-(n-1)/2} x_2 (x_1)^{n-1} + 
q_S^{-(n-3)/2} x_1 x_2 (x_1)^{n-2} + ...+ q_S^{(n-1)/2} (x_1)^{n-1} x_2 
      \nonumber\\
  &=& q_S^{-(n-1)/2}(1 + q_S^2 + ... + q_S^{2(n-1)}) x_2 (x_1)^{n-1} 
      \nonumber\\
  &=& q_S^{(n-1)/2} [n]_{q_S}  x_2 (x_1)^{n-1}.
\label{Xm_calc}
\err
Since $[k M_S]_{q_S} =0$, it follows that $X_i^- \cdot (x_1)^{k M_S} =0$ 
for all $i$, hence $(x_1)^{k M_S}$ is a one--dimensional \rep
of $\U^{fin}$. As discussed in Section \ref{subsec:reps} 
below (\ref{X_i_0_L}),
this implies that all weights of $\F(k M_S)$ have the form 
$\l_z = \sum z_i M_i \L_i$, and $\F(k M_S)$ is a \rep of the classical
universal enveloping algebra $U(\tilde \mg)$ with generators
$\tilde X_i^{\pm}$ and $\tilde H_i$. Moreover, it is a highest weight
module with highest weight vector $(x_1)^{k M_S}$. By the 
classical \rep theory, it follows that it is irreducible, hence
\be
\F(k M_S) \cong L^{res}(k M_S \l_1),
\ee
which is essentially  $L(k \l_1)$ of $\tilde \mg$ (except for $D=3$, where
it is $L(k \L_1)$).

Now consider more generally $n = n_0 + k M_S$ with 
$0 \leq n_0 < k_S$ and $k \in \N$. Then both $\F(n)$ 
and $\F(n_0) \tens \F(k M_S)$ are highest weight modules with 
highest weight $n \l_1$. Clearly $\F(n) \subset \F(n_0) \F(k M_S)$,
by (\ref{F_k}) and (\ref{coproduct_X}). On the other hand,
$\F(n_0) \F(k M_S) \subset \F(n_0) \tens \F(k M_S)$, which is 
isomorphic to $L^{res}(n \l_1)$ by Theorem \ref{L_res_theorem}.
Hence we have shown that 

\begin{theorem}
For $n = n_0 + k M_S$ with $0 \leq n_0 < k_S$ and $k \in \N$,
\be
\F(n) = \F(n_0) \F(k M_S) \cong \F(n_0) \tens \F(k M_S) \cong 
  L^{res}(n \l_1).
\label{V_decomp}
\ee
\label{V_theorem}
\end{theorem}
In particular, it is essentially the tensor product of $\F(n_0)$
with the irreducible \rep $L(k \l_1)$ (or $L(k \L_1)$ for $D=3$) 
of the classical algebra $\tilde \mg$.

\sect{Real forms and Hilbert space \reps of the quantum spaces}
\label{sec:hilbert}

\subsection{Hilbert spaces for $S_q^{D-1}$ and $AdS_q^{D-1}$}
\label{subsec:hilbert_spaces}

Now we are ready to discuss the reality structure
at roots of unity. As was pointed out before, 
(\ref{euclid_star}) and (\ref{ads_real}) are not consistent 
with the algebra for $q$ a phase. To find the correct definition, we 
first construct the Hilbert spaces, and then simply calculate the adjoint of
the operators of interest. We want to emphasize that the 
inner products on irreducible \reps are determined uniquely by
(\ref{invariant}). Indeed on any highest weight module, 
there exists a unique (up to normalization) invariant inner product 
for a given star structure of the form (\ref{first}).
This is because once the inner product is defined on the highest weight vector,
it can be calculated for all descendant vectors using 
(\ref{invariant}); in general, it is not unitary.
The resulting invariant inner product is non--degenerate
if the \rep is irreducible.

We first discuss the quantum sphere.
As \rep of $\U^{res}$, we can consider 
\be
\F(n) =  \U^{res} \cdot (t_1)^n \subset S_{q,\C}^{D-1}
\ee
instead of (\ref{F_k}).
However, not all these $\F(n)$ should be considered as fields on
the ``real'' quantum sphere, only those which are unitary \reps of the
compact form 
\be
{X_i^{\pm}}^* = X_i^{\mp}, \quad H_i^* = H_i,
\ee
of $U_q^{fin}(so(D))$, with the natural inner product discussed above. 
It was proved in \cite{nc_reps} that this certainly holds for
those $L^{res}(k \l_1) = L^{fin}(k \l_1)$ with $k \leq k_S$ (\ref{k_S}).
Here the assumption (\ref{q_root}) is important. 

Hence we could define a Hilbert space of 
functions on the real quantum sphere to be the direct sum
of all $\F(k)$ with $k \leq k_S$. In order to obtain a simple definition of
position operators in Section \ref{subsec:pos_ops}, 
we impose the slightly stronger bound $k < k_S$, and define the 
Hilbert space of functions on the {\em real quantum sphere} to be
\be
S_q^{D-1}:= \bigoplus\limits_{0 \leq n_0 < k_S}  \;\; \F(n_0) = 
            \bigoplus_{0 \leq n_0 < k_S} \;\; \U^{fin} \cdot (t_1)^{n_0}.
\label{real_sphere}
\ee
The position operators will be discussed in Section \ref{subsec:pos_ops}.
Its generators are essentially the $t_i$ as in Section 2; they will have 
to be slightly modified in order to account for the cutoff. Their
star structure is determined uniquely by the adjoint on the Hilbert space, 
and will be given explicitly in Section \ref{subsec:pos_ops}.
$S_q^{D-1}$ is a truncated version of the 
classical sphere $S^{D-1} = \oplus_{n \geq 0} L(n \l_1)$,
which is recovered in the limit 
$M\rightarrow \infty$, or $q \rightarrow 1$. 

Now consider more generally $n=n_0 + k M_S$ with $0 \leq n_0 < k_S$, 
so that $\F(n) = \F(n_0) \tens \F(k M_S)$ according to (\ref{V_decomp}). 
Since $\F(k M_S) \cong L^{res}(k M_S\l_1)$ 
is a finite--dimensional irreducible \rep of the classical $\tilde \mg$,
it has a standard positive--definite invariant inner product.
We just discussed the inner product on $\F(n_0)$. 
This suggests a natural positive definite inner product on 
$\F(n)$ as the tensor product, so that 
\be
(f \rho_{(k)}, f' \rho'_{(k)}):= (f,f')(\rho_{(k)},\rho'_{(k)})
\label{inner_F}
\ee 
where $f,f' \in \F(n_0)$ and $\rho_{(k)},\rho'_{(k)} \in  \F(k M_S)$.
This leads to an interesting physical interpretation, as we will see. 
We will always use an orthonormal basis corresponding to
this inner product from now on.

We have seen in the previous section
that all weights of $L^{res}(k M_S \l_1)$ have the form
$\l_{z} = \sum_i z_i M_i \L_i$. Therefore $\F(n)$ is the 
direct sum of irreducible \reps 
$L^{fin}(n_0\l_1 + \l_{z})$ of $\U^{fin}$ for various $\l_z$; if the
multiplicity of a certain $L^{fin}(n_0\l_1 + \l_{z})$ in $\F(n)$ is larger
than one, then the corresponding subspace can be decomposed as 
an orthogonal sum of irreducible components. Hence 
$\F(n)$ is a direct orthogonal sum of Hilbert spaces
of type $L^{fin}(n_0\l_1 + \l_{z})$, with the induced inner 
product (\ref{inner_F}), and the different components are
related by the action of the classical generators 
$\tilde X_i^{\pm}$ (\ref{class_generators}).
One can now calculate the adjoint of the generators of $\U^{fin}$
on $L^{fin}(n_0\l_1 + \l_{z})$ with respect to this inner product.
The result is (see \cite{nc_reps}, Theorem 5.1)
\be
H_i^* = H_i, \quad {X_i^{\pm}}^* = s_i X_i^{\mp}, \quad \mbox{where}
         \quad s_i = (-1)^{z_i}.
\label{real_U}
\ee
If $\pi^j_i(u)$ is the \rep of $u \in \U^{fin}$ 
on $L^{fin}(n_0\l_1 + \l_{z})$ 
with respect to an orthonormal basis, this means that
\be
(\pi(u)^{\dagger})^j_i = \pi^i_j(u)^* = \pi^j_i(u^*)
\label{unitary_basis1}
\ee
as in (\ref{unitary_basis}); here $*$ really depends on $z$ as in
(\ref{real_U}), which will however be suppressed in the following.
Hence $L^{fin}(n_0\l_1 + \l_{z})$ is a unitary \rep of a certain real 
form of the type (\ref{first}) of $U_q^{fin}(so(D,\C))$, 
and the ``sectors'' with different $\l_{z}$ but the same $s_i$
are unitary with respect to the same real form form.
Comparing with Section \ref{subsec:star_U}, we conclude that
if all $z_i$ are even, then
$L^{fin}(n_0 \l_1 + \l_{z})$ is a unitary \rep of the 
compact form (\ref{euclid_involution}) and hence a scalar field
on $S_q^{D-1}$. If the $z_i$ are such that $s_i = (-1)^{z_i}$ is as in 
(\ref{ads_involution}), then it is a unitary \rep of the Anti--de Sitter group 
$\U^{fin}(so(2,D-2))$, and should in fact be viewed as a scalar field on
$AdS_q^{D-1}$,  corresponding to a square--integrable scalar field
in the classical case. To understand this, 
consider the $L^{fin}((2M_S-k)\l_1)$ as lowest--weight \reps 
$L_{fin}(k \l_1)$
of $\U^{fin}$ with lowest weight $k \l_1$. For low energies, they have 
the same structure as the scalar fields 
on the classical AdS space, which are irreducible unitary lowest--weight 
representations of $SO(2,D-2)$ 
realized in terms of  square--integrable functions.
It is very remarkable that they are realized here 
in terms of polynomials in the coordinate functions.
Therefore we define the Hilbert space
of functions on the {\em real quantum Anti--de Sitter space} to be
\be
AdS_q^{D-1}:= \bigoplus_{M_S \leq n < M_S + k_S}\;\; \U^{fin} \cdot (t_1)^n
    = \bigoplus_{M_S \leq n < M_S + k_S}\;\; L^{fin}(n \l_1)
    = S_q^{D-1} (t_1)^{M_S}.
\label{real_AdS}
\ee
In terms of lowest--weight \reps 
$L_{fin}(n \l_1) = \U^{fin} \cdot (t_D^{M_S-n} t_1^{M_S})$
of $\U^{fin}$, this can be written as 
\be
AdS_q^{D-1} = \bigoplus_{(D-2)/2 < n \leq M_S}\;\; L_{fin}(n \l_1).
\ee
For energies less than $M_S$,
the states in the Hilbert spaces are the same as classically,
and the action of $\U^{fin}$ approaches the classical one for any given weight
as $M\rightarrow \infty$. The energy of all states is less than $2 M_S$.  
The precise definition of the 
position operators will be given in Section \ref{subsec:pos_ops}.
In the classical case, the lower bound $(D-2)/2$ 
can be seen by a simple
dimensional argument; however it can be violated slightly.

For even $D>4$, the irreducible quotient of
$\U^{fin}\cdot (t_D^{M_S-(D-4)/2} t_1^{M_S})$ is the scalar singleton
$L_{fin}((D-4)/2 \l_1)$, with lowest weight $(D-4)/2 \;\l_1$; 
we will not consider it any more here.

\subsection{Product spaces}
\label{subsec:product_spaces}

From the above discussion, it would seem much more natural to consider
all polynomials in the $t_i$ instead of just certain $\U^{fin}\cdot (t_1)^n$.
For simplicity, we will restrict ourselves to the polynomials of the form
$\F(n) = \U^{res} \cdot (t_1)^n$ for all $k M_S < n < k M_S + k_S$
and all $k \in \N$, and study their field content.
Let $\widetilde{\l_1} = \l_1$ for $D > 3$, and 
$\widetilde{\l_1}= \l_1/2 = \L_1$  for $D=3$.
Then $\F(n) = \F(n_0) \F(k M_S)$ using (\ref{V_decomp}), 
where the second factor is essentially the \rep $L(k \tilde{\l_1})$
of the classical symmetry algebra $\tilde \mg$, which
connects the various components with different $\l_{z'}$. Hence 
\be 
\oplus \F(n) = \Big(\bigoplus_{0 \leq n_0 < k_S} \F(n_0)\Big)\; \tens \;
            \Big( \bigoplus_{k \in \N} L(k \widetilde{\l_1}) \Big),
\ee
where certain modes were omitted as in Section \ref{subsec:hilbert_spaces}
for simplicity. For the moment we ignore the reality structure.

Observe that the $L(k \widetilde{\l_1})$ are very particular representations 
of the classical $\tilde \mg$, which have a nice interpretation.
Consider first $D=2r$. Then the dual algebra is $\widetilde{so}(2r) = so(2r)$ 
as shown in \cite{nc_reps}, and $L(k \widetilde{\l_1})$  can be viewed
as a function on the classical sphere $S^{D-1}$. Hence
$\oplus_{k \in \N} L(k \widetilde{\l_1}) \cong Fun(S^{D-1})$,
which is the space of polynomial functions on the classical sphere $S^{D-1}$. 

Next, consider $D=2r+1$, which is less obvious. Then $\widetilde{\l_1} = \L_1$, 
and $\F(k M_S)$ is the highest--weight \rep $L(k \L_1)$ of 
$\widetilde{so}(2r+1) = sp(2r)$ with highest weight $k \L_1$. Observe that
the $2r$--dimensional \rep $L(\L_1)$ is not real, 
in the sense that the $2r$  variables $z_i$ are necessarily complex,
and can be considered as $4r$ real variables $x_i$. The
compact form $USp(2r)$ acts by multiplying the $z_i$ with a unitary 
matrix. Therefore
the radius $x^2 = \sum_i z_i \obar{z_i}$ is invariant 
under $USp(2r)$, and the orbit of $USp(2r)$ in $L(\L_1) = \C^{2r}$ 
turns out to be sphere $S^{4r-1}$ (to see this, one can show that the 
stabilizer of $(1,0,...,0)$ is $USp(2r-2)$, hence 
$USp(2r)/ USp(2r-2) \cong S^{4r-1}$ on dimensional grounds).
The set of polynomials of degree $k$ in the $z_i$ is precisely the \rep  
$L(k \L_1)$, hence $\oplus_{k \in \N} L(k \L_1)$ is the space of 
those functions on $S^{4r-1}$ which are induced by holomorphic functions 
on $\C^{2r}$; for lack of a better name, we call it the ``chiral sector'' 
$S_{\chi}^{4r-1}$ of $S^{4r-1}$. The full sphere could be obtained 
by tensoring $S_{\chi}$ with another (conjugate) chiral sphere.

In summary, we found (ignoring the reality structure)
$\oplus_{n \in \N} \F(n) \cong S_{q}^{D-1} \tens Fun(M)$, where
\be
M = \left\{ \begin{array}{rl} S^{D-1}, & D-1 \;\;\mbox{odd} \\
                              S_{\chi}^{2D-3}, & D-1 \;\;\mbox{even},
            \end{array} \right. 
\ee
which should be interpreted as the functions on the product of the 
quantum sphere with the classical sphere, or the ``chiral sector'' of 
the classical sphere.

Now we take into account the star structure (\ref{real_U}) induced on 
the various $L^{fin}(n_0\l_1 + \l_{z'}) \subset \F(n)$. 
As we have seen before, certain sectors 
$\l_{z'}$ of the modules $\F(n)$ correspond to $S_q^{D-1}$, others
correspond to $AdS_q^{D-1}$, and others yet have not been identified here.
This means that we obtain the product of $AdS_q^{D-1}$ or $S_q^{D-1}$ 
with certain quotients (=orbifolds) of the classical spaces.
The functions on the orbifolds are described by functions on the 
covering space which are invariant under the action 
of a certain classical group, possibly modulo signs (for twisted orbifolds). 

To make this more specific, recall that
$v_{z'} \in L^{fin}(n_0\l_1 + \l_{z'}) \subset \F(n)$ is 
an element of $AdS_q^{D-1}$ if
$s_j = (-1)^{z'_j}$ is as stated in (\ref{ads_involution}).
By Theorem \ref{classical_algebra}, this condition can be restated as 
\be
e^{i \pi \tilde H_j}\cdot v_{z'} = s_j v_{z'}.
\ee
Here the $e^{i \pi \tilde H_j}$ are elements of the classical group
corresponding to $\tilde \mg$, whose square is the identity;
they can be viewed as rotations by $\pi$. They generate the abelian group
\be
\Gamma = (Z_2)^r.
\ee
This means that the classical functions $v_{z'}$ 
on $M = S^{D-1}$ or $S_{\chi}^{2D-3}$ are really functions on the (twisted)
orbifold $M/\Gamma$. Hence 
\be
\oplus \F(n) = 
\Big( AdS_q^{D-1}\times M/\Gamma \Big)  \;\; \oplus \;\;  
\Big( S_q^{D-1} \times M/\Gamma \Big)  \;\; \oplus \dots
\ee
in somewhat sloppy notation, where the different summands correspond
to different twisted sectors of the orbifold $M / \Gamma$.
In fact, some of these other sectors (which were not considered so far)
are again equivalent to $AdS_q^{D-1}$. For example for $D=5$, all
sectors turn out to be either $S_q^{4}$ or $AdS_q^4$. Hence they might 
recombine, and $(Z_2)^r$ might effectively be smaller. 

The residual symmetry of $\tilde \mg$ on these orbifolds is \cite{nc_reps}
$su(2)^r$ for $D=2r+1$, and to $u(1)^r$ for $D=2r$; in a
sense, $\tilde \mg$ is spontaneously broken.

\subsection{The Universal Weyl Element $\w$}
\label{subsec:weyl}

The appropriate mathematical tool to describe the involution is an element 
of an extension of $\U$ by generators $\w_i$ of the braid group, introduced
in \cite{kirill_resh} and \cite{lev_soib}. The $\w_i$ act on \reps of 
$\U^{res}$, and define a braid group action  
on $\U$ via $T_i(u) = \w_i u \w_i^{-1}$ for $u\in \U$.
All we need is the generator corresponding to the longest element
of the Weyl group, $\w$. Acting on an irreducible representation, $\w$ maps
the highest weight vector 
into the lowest weight vector of the contragredient (dual) representation,
as classically.
It has the following important properties \cite{kirill_resh,lev_soib}:
\berr
\Del(\w) &=& \RR^{-1}\w\tens \w = \w\tens \w \RR^{-1}_{21}, \label{del_om} \\
\w u \w^{-1} &=& \theta S \g(u) = S^{-1} \theta \g(u),  \label{autom_w} \\
\w^2 &=& v \epsilon,                               \label{w_2}
\err
for $u \in \U$,
where $\g = \id$ in the case $so(2r+1)$, and $\g$  
is the automorphism of the Dynkin diagram\footnote{for $so(8)$,
it is the automorphism which interchanges $\a_3$ with $\a_4$
and leaves $\L_1$ invariant, as can be seen from $V_D$.}
in the case $so(2r)$.  $\epsilon$ is a Casimir with 
\be
\Del(\epsilon) = \epsilon\tens\epsilon
\label{del_eps}
\ee
which takes the values $\pm 1$. It turns out that 
$\epsilon = -1$ only for spinorial \reps which are not considered in this 
paper, thus we will put $\epsilon =1$ from now on.
(\ref{del_om}) justifies the name "universal Weyl element" for $\w$.
One can also define antipode and counit for $\w$, and
explicit formulas for $\w$ in terms of formal sums in $\U$ can be given
\cite{jantzen}. Since explicit proofs of (\ref{autom_w}) and (\ref{w_2}) 
have only been given for the rank one case in the literature, 
we will sketch short proofs for the general case in Appendix B.

For our purpose, we define a slightly modified element 
\be
\tw := q^{2\tilde{\rho}} \w,
\label{tw} 
\ee
with the same properties as $\w$ except
\be
\tw u \tw^{-1} = \theta S^{-1} \g(u) = S \theta \g(u).     \label{autom_tw}
\ee
$\w$ or $\tw$ has a simple interpretation for real representations.
A \rep $V$ of $\U^{res}$ will be called {\em real} here if there
exists an intertwiner
\be
g: V \tens V \rightarrow \C,
\ee
or in other words an invariant tensor on $V \tens V$.
If $g_{ij}$ is the matrix of this intertwiner, then
\be
g_{ij} \pi^i_k(u_1) \pi^j_l(u_2) = \eps(u) g_{kl}, 
\label{g_invar}
\ee
for all $u \in \U^{res}$,
which by standard Hopf algebra identities is equivalent to 
\be
g_{ij} \pi^i_k(u) = g_{kl} \pi(S(u))^l_j,
\ee
or $\pi(u)^T g = g \pi(S(u))$ in matrix language. Now by (\ref{autom_tw}),
one can write $S(u) =  \tw \theta(\g(u)) \tw^{-1}$,
and one obtains
\be
\pi(u)^T = A^{-1} \pi(\theta(\g(u))) A,
\label{A_eqn}
\ee
where $A^{-1} = g \pi(\tw)$. 

In the classical case, it is known that  
the ``real'' \reps of a Lie group can be chosen to be 
orthogonal or symplectic matrices, in a suitable basis.
Moreover, a highest weight \rep is real if the highest weight is invariant
under the automorphism $\g$.
We can show the following related statement in the $q$--deformed case:

\begin{lemma}
If $\g$ acts trivially on the dominant integral weight
$\l$, i.e. $<H_i,\l> = \newline <\g(H_i), \l>$ for all $i$, then 
the automorphism $\g$ can be realized as a conjugation on $L^{res}(\l)$,
\be
\pi(\g(u)) = C \pi(u) C                    \label{C_conj}
\ee
with $C^2 = \one$, and $L^{res}(\l)$ is real.
If $L^{res}(\l)$ is in addition a unitary \rep of the compact form 
(\ref{euclid_involution}) of $\U^{res}$, then
there exists an orthonormal basis of $L^{res}(\l)$ 
such that the matrices $\pi(u)^i_j$ satisfy 
\be 
\pi(u)^T =  J^{-1} \pi(\theta(u)) J
\label{orthogonal_basis}
\ee
and (\ref{unitary_basis1}) for all $u \in \U^{res}$,
where $J = \one$ if $L(\l)$ is orthogonal at $q=1$,
and $J$ is block--diagonal with blocks of the form 
$\(\begin{array}{cc} 0 & -1 \\ 1 & 0 \end{array} \)$
if $L(\l)$ is symplectic at $q=1$. In this basis, $C^{\dagger} = C$ and
$C^T = J C J^{-1}$.
\label{orth_lemma}
\end{lemma}
This is proved in Appendix C. An alternative approach has been given 
in \cite{thesis}. 
We will always use such a basis from now on. Even though 
these properties will be used only for the vector \rep $V_D$ in this paper 
where they can be checked directly, they hold much more generally.
In particular, (\ref{orthogonal_basis}) together with (\ref{A_eqn})
implies by Schur's Lemma that the invariant tensor is given by
\be
\hat g = J^T C \pi(\tw^{-1}),                      \label{g_JC}
\ee
where the hat on $\hat g$ will indicate the particular normalization implied by
this equation.
Further consistency relations of the $\hat R$--matrix with the
invariant tensor follow from (\ref{del_om}), see \cite{resh_1} or
\cite{thesis}.

Now we concentrate on the case of interest here, which is the
vector \rep $V_D \cong L^{res}(\l_1)$. 
Since $\g$ acts trivially on the highest weight, $V_D$ is real,
and the invariant tensor $g$ is nothing but the metric (\ref{metric}).
Moreover $V_D$ is orthogonal classically, hence $J=1$, and 
\be
\pi(u)^T =  \pi(\theta(u)) 
\label{orthogonal_basis1}
\ee
for all $u \in \U^{res}$. Therefore $\hat g_{ij} = C^i_k \pi^k_j(\tw^{-1})$, 
and $C^i_j = \d^i_j$ for $so(2n+1)$.

Finally, the star structure (\ref{euclid_involution}) on
$\U$ can be extended by $\w^* = \w^{-1}$. This is 
consistent with (\ref{del_om}), (\ref{autom_w}) and (\ref{w_2})
for all real forms (\ref{real_U}) of $\U$,
and also with the explicit formulas for $\w$. 
Acting with (\ref{autom_w}) on a unitary irreducible representation,
it follows by Schur's Lemma that $\pi(\w)^{\dagger} \pi(\w) = 1$, 
hence $\pi(\w)^{\dagger} = \pi(\w^*)$, as it should be.
By (\ref{g_JC}), this implies that
\be
\hat g^{\dagger} = \hat g^{-1}.
\label{g_dagger}
\ee
For $V_D$, this means  $\hat g_{ij}^* = \hat g^{ji}$,
which is well-known.

\subsection{Pre--involution on the extended coordinate algebras}
\label{subsec:invol_ext}

In this section, we construct an auxiliary ``pre--involution'' on certain 
algebras of coordinate functions extended by $\U^{res}$, which will
be used in the next section to write down 
the star structure on the (extended) coordinate algebra.

Let $\F$ be a $\U^{res}$--module algebra, which in the present paper 
will be either $\C_q^D$ or $S_{q,\C}^{D-1}$. The following constructions 
are however much more general.

It turns out one has to extend $\F$ in order to write down the involution.
Consider the {\em semi--direct product (cross product) algebra} 
$\U^{res}\smash \F$. As a vector space, this is nothing but 
$\U^{res}\tens \F$, with the relations 
\be
u x_i = (u_1 \cdot x_i) u_2
\label{smash}
\ee
for $x_i \in \F$ and $\Delta(u) = u_1 \tens u_2$ for $u \in \U$. 
This defines a consistent algebra, because $\U^{res}$ 
is a Hopf algebra and $\F$ is a (left) $\U^{res}$ --module algebra. 

Let $\hat g^{ij}$ be the inverse of the metric,
i.e. $\hat g^{ij} \hat g_{jk} = \d^i_k$, and define as usual \cite{FRT}
\be
{L^-}^i_j = \pi^i_j(\RR_1^{-1}) \RR_2^{-1}.             \label{L_-}
\ee
Then we can define an auxiliary ``pre--involution''
on $\U^{res}\smash \F$ as follows.

\begin{prop}
Let $\F$ be $\C_q^D$ or $S_{q,\C}^{D-1}$, with generators  $x_i$.
The map $\obar{(.)}$ on $\U^{res}\smash \F$ defined by 
\berr
\obar{x_i} &=& \tw x_j C^j_i \tw^{-1} = x_l {L^-}^l_k \hat g^{ki} \nn\\
\obar{H_i} &=& H_i, \;\; \obar{X_i^{\pm}} = X_i^{\mp}   \label{invol_smash}
\err
can be consistently 
extended as an antilinear anti--algebra map on  $\U^{res}\smash \F$,
which satisfies
\be
\obar{\obar{x}} = v x v^{-1}              \label{pre_invol}
\ee
for any $x \in \U^{res}\smash \F$. Moreover,
\be
\obar{x^2} = \obar{x_i x_j g^{ij}} = x^2,
\label{r_2_reality}
\ee
if the normalization of $g^{ij} = q_S^{(D-1)/2} \hat g^{ij}$ is 
such that $g^{ij} = g_{ij}$.
\label{involution_theorem}
\end{prop}

On $\U^{res}$, this is just the compact involution 
considered before; notice that $v$ is central in $\U^{res}$.
For $q=1$, ${L^-}^i_j = \d^i_j$, and
(\ref{invol_smash}) agrees with (\ref{euclid_star})
in the classical limit. In  particular, the use of $\tw$ is not essential,
but a considerable simplification. 

This generalizes easily for more general algebras generated by 
generators $x_i$ in bigger, real representations.
The $\RR$--matrix involved in $\obar{x_i}$ essentially
corrects the flip in the conjugation of the coproduct of $\U$
which occurs if $q$ is a phase. In the form with $L^-$, 
a somewhat similar formula was proposed in \cite{mack_schom} 
in a different context. However, we use it as an
intermediate step towards the correct ``physical'' involution.
The two formulas for $\obar{x_i}$ are identical, because
\berr
\tw x_j C^j_i \tw^{-1} 
      &=&  x_l \pi^l_k(\RR_1^{-1}) \pi^k_j(\tw) C^j_i \RR_2^{-1} \nn\\
      &=& x_l {L^-}^l_k \hat g^{ki} 
\label{invol_explic}
\err
where we used (\ref{del_om}) and
$\pi^k_j(\tw) C^j_i = (\hat g^{-1})^k_i = \hat g^{ki}$, since $J=1$ here.

\begin{proof}
One has to verify that the map is consistent with (\ref{smash}); 
this is done in Appendix C. 
The verification that $\obar{\obar{x_i}} = v x_i v^{-1}$ is immediate
if one uses $\obar{\tw}  = \tw^{-1}$ and $C^* = C^{\dagger T} = C$.
Recall that we do not consider
spinorial \reps here, so $\epsilon =1$ in (\ref{w_2}).
In the more ``down--to--earth'' version with $L^-$, the calculation
is more complicated, and sketched in the Appendix C.

The bar is compatible with $(P^-)^{ij}_{kl} x_i x_j =0$, using the
fact that $R^* = R^{-1}$; note that for even $D$, 
$\hat R$ commutes with $C \tens C$, because $\RR$ is invariant under 
the automorphism $\g$.

To see the last relation, observe that $C^i_k C^j_l g_{ij}$ 
is also an invariant tensor, hence it is proportonal to $g_{kl}$,
and in fact it is equal to $g_{kl}$
(consider e.g. the highest weight state, where $C=1$). By (\ref{g_JC}), this
implies that $C \pi(\tw) = \pi(\tw) C$ and therefore
$g^2 = q_S^{-(D-1)} \pi(v^{-1}) = \one$, so that $g^{-1}$ is equal to $g$.
This is known explicitly for $V_D$. Hence (\ref{r_2_reality}) 
follows from $\tw x^2 \tw^{-1} = x^2$, since $x^2$ is a singlet.
Therefore the bar is well--defined on $S_{q,\C}^{D-1}$.
\end{proof}


\subsection{Position operators and star structure}
\label{subsec:pos_ops}

We concentrate on $S_{q,\C}^{D-1}$ from now on. The generators $t_i$ 
define linear operators on $S_{q,\C}^{D-1}$, and we now restrict them
to the real sphere $S_q^{D-1} = \oplus_{0 \leq n_0 < k_S}  \;\; \F(n_0)$.
Since $t_i \F(k) \in \F(k+1) \oplus \F(k-1)$ and using the fact that
all $\F(k)$ for $0 \leq k < k_S+1$ are linearly independent by Theorem 
\ref{M_k_decomp}, the projector
\be
P_c: \bigoplus_{0 \leq n_0 < k_S+1}  \;\; \F(n_0) \rightarrow 
     \bigoplus_{0 \leq n_0 < k_S}  \;\; \F(n_0) = S_q^{D-1}
\ee
is well--defined\footnote{this is no longer true if we include more modes in 
$S_q^{D-1}$ than in (\ref{real_sphere}), since then non--decomposable
modules appear.}, commutes with $\U^{res}$, and 
maps $t_i S_q^{D-1} \subset S_{q,\C}^{D-1}$ into $S_q^{D-1}$. 
Hence the operator
\be
\hat t_i := P_c t_i P_c
\label{t_hat}
\ee
acts on $S_{q}^{D-1}$, and generates the algebra of functions on 
the real quantum sphere. Of course for $k < k_S-1$, it coincides
with $t_i$. Hence from now on, we can consider $S_q^{D-1}$ as
an algebra, generated by the $\hat t_i$. As before,
it can be extended to $\U^{res}\smash S_{q}^{D-1}$,
which we consider as an {\em operator algebra} acting on $S_q^{D-1}$.
This will be understood from now on.
Since $S_q^{D-1}$ is a Hilbert space, we can calculate the star on that 
algebra, which is given by the adjoint. We will do this by constructing
the inner product on the Hilbert space explicitly.

Observe that the ``pre--involution'' defined  in 
Proposition \ref{involution_theorem} induces a ``pre--involution''
on $\U^{res}\smash S_{q}^{D-1}$ by 
\be
\obar{\hat t_i} = \tw \hat t_j C^j_i \tw^{-1},
\ee
since
\berr
P_c \obar{t_i} P_c &=& P_c \tw t_j C^j_i \tw^{-1} P_c = 
       \tw P_c t_j P_c C^j_i \tw^{-1} \nn\\
   &=& \tw \hat t_j C^j_i \tw^{-1} = \obar{\hat t_i}
\err
if acting on $S_q^{D-1}$. Moreover, there exists a unique invariant state
$<\;>$ on $S_{q}^{D-1}$, given by the projection on the
invariant component (singlet) in the decomposition 
(\ref{real_sphere}), i.e.
\be
<f(\hat t)> = f_0 \in \C,
\label{invar_state}
\ee
where $f_0$ is the unique trivial component of $f(\hat t)\in S_{q}^{D-1}$. 
Because it is invariant, i.e. 
$<u\cdot f> = \eps(u) <f>$, this state extends naturally to a state on 
$\U^{res}\smash S_{q}^{D-1}$, so that $<u f> = \eps(u) <f> = <f u>$.
It satisfies
\be
<f>^* = <\obar{f}>                            \label{state_bar}
\ee
for $f\in S_{q}^{D-1}$, which again extends to 
$f \in \U^{res}\smash S_{q}^{D-1}$. 
This can be used to define an inner product $(\;,\;)_0$
on $S_{q}^{D-1}$, 
\be
(f,g)_0:=<\obar{f}g>,
\ee
which however is not hermitian, rather 
\be
(f,g)_0^* = <\obar{\obar{f}g}> = <\obar{g}vfv^{-1}> = (g,v\cdot f)_0.
\label{herm_0}
\ee
It satisfies furthermore
\be
(f,u\cdot g)_0 = <\obar{f}ug> = <\obar{u^* f}g> = (u^*\cdot f,g)_0
\label{u_star_0}
\ee
where $u^*$ is the compact involution (\ref{euclid_involution}),
by (\ref{invol_smash}). 

Now it is easy to write down an invariant hermitian inner product 
on $S_{q}^{D-1}$.
We need the square--root of the Casimir $v$ (\ref{v})
acting on the \rep $S_{q}^{D-1}$. 
Since $v$ is diagonal on $S_{q}^{D-1}$ with eigenvalues 
$q^{-c_{\l}}$ by Theorem \ref{scalar_thm}, 
we can simply define the operator $\sqrt{v}$ acting on $S_q^{D-1}$ by
\be
\sqrt{v}:= q^{-\frac 12 c_{n \l_1}} \quad \mbox{on } \F(n),
\label{sqrt_v}
\ee 
and similarly $\sqrt{v^{-1}}$. Here $\U^{res}$ is considered as an
algebra of operators rather than an abstract (Hopf) algebra, 
hence there is no problem adding this new operator $\sqrt{v}$.
One may also give a more formal definition in terms of a rational
function of $v$, see \cite{thesis}.
This allows to define 
\be
\Om := \tw \sqrt{v^{-1}},
\label{Omega}
\ee
again as an operator on $S_{q}^{D-1}$ (or similar spaces).
It satisfies $\Om u \Om = \theta S^{-1} \g(u) = S \theta \g(u)$, and
\be 
\Om^2 =1
\ee 
if acting on $S_{q}^{D-1}$, by (\ref{w_2}). 
Then the following holds:

\begin{theorem} (``Quantum Euclidean sphere'')
For any $f,g \in S_{q}^{D-1}$, 
\be
(f,g):=(f,\sqrt{v}\cdot g)_0 
\label{inner_S}
\ee
defines a positive definite hermitian inner product on $S_{q}^{D-1}$.
It satisfies
\berr
(u \cdot f,g) &=& (f,u^*\cdot g)      \label{u_star}\\
(\hat t_i f,g)     &=& (f,\hat t_i^* g)         \label{t_i_star}
\err
where $u^*$ is the compact involution (\ref{euclid_involution}) on $\U^{res}$,
and 
\be
\hat t_i^* = \Om \hat t_j C_j^i \Om = \
             \sqrt{v^{-1}}\; \hat t_j (L^-)^j_k \; \hat g^{ki} \sqrt{v}
\label{t_hat_adjoint_S}
\ee
as operators on $S_{q}^{D-1}$.
\label{invol_thm}
\end{theorem}

Of course, 
$(f,\hat t_i g) = (\hat t_i^* f,g)$
also holds by hermiticity. In particular, this defines a star structure 
on the {\em operator} algebra generated by $\hat t_i$ and $\U^{res}$ 
acting on $S_{q}^{D-1}$. 

\begin{proof}
(\ref{t_i_star}) follows from
\berr
(\hat t_i f,g) &=& (\hat t_i f, \sqrt{v}\cdot g)_0 \nn\\
     &=& <\obar{f} \obar{\hat t_i} \sqrt{v} \cdot f> \nn\\ 
   &=& <\obar{f} \tw \hat t_j C^j_i \tw^{-1} \sqrt{v} \cdot g> \nn\\
    &=& <\obar{f} \sqrt{v} \Om \hat t_j C^j_i \Om^{-1} g> \nn\\
    &=& (f,\Om \hat t_j C^j_i \Om^{-1} g).
\err
(\ref{u_star}) follows immediately from (\ref{u_star_0}), and 
hermiticity is seen easily using (\ref{herm_0}):
\berr
(f,g)^* &=& (f, \sqrt{v} \cdot g)_0^*  = (\sqrt{v^{-1}}\cdot f,g)_0^*  \nn\\
        &=& (g, v \sqrt{v^{-1}} \cdot f)_0 \nn\\
        &=& (g,f).
\err
It only remains to prove that it is positive definite. For this,
it is enough to check that $((t_1)^n,(t_1)^n) >0$ for all $n < k_S$,
using the invariance of $(\;,\;)$  and unitarity of $L^{res}(n \l_1)$ for 
$n < k_S$. This follows either by a continuity
argument for $q'$ on the unit circle between 1 and $q$ 
using (\ref{q_root}), or by a direct calculation, 
which is done in Appendix C. There, the bound $n < k_S$ will seen 
explicitly (the reason is that for larger $n$, $t_i\F(n)$ becomes
indecomposable).
\end{proof}

This completes the definition of 
the real quantum sphere $S_q^{D-1}$. Together with $\Om$ or $\U^{res}$, 
it forms a star algebra. In the classical limit $q \rightarrow 1$,
the star becomes the correct involution 
$t_i^* = t_j g^{ji}$ (\ref{euclid_star}) for any given mode $\F(n)$, 
as was pointed out below Proposition \ref{involution_theorem}.  
Moreover, $S_q^{D-1} \subset S_{q'}^{D-1}$ naturally as a vector space if 
$q' = e^{i \pi/M'}$ with $M' > M$, since the 
irreducible \reps $L^{res}(n\l_1)$ of $\U^{res}$ and ${\cal U}_{q'}^{res}$ 
with degrees $n \leq  k_S$ are isomorphic.
Real, self--adjoint position operators can be defined for example as
$\hat t_i + \hat t_i^*$. This is similar to the ``fuzzy sphere'' 
\cite{madore} in the sense that there is a cutoff in the set of \reps
of the symmetry group, but we do not insist here that the algebra of 
functions should be a simple matrix algebra; instead, we consider a
larger algebra including $\Om$ or $\U$.

Finally we define the algebra of coordinate functions or position 
operators on the real Anti--de Sitter space $AdS_q^{D-1}$, 
which was our main goal. Since the Hilbert space of scalar fields 
(\ref{real_AdS}) (corresponding to square--integrable functions) 
has the structure 
\be
AdS_q^{D-1} = S_q^{D-1} \tens L^{fin}(M_S \l_1) = S_q^{D-1} \rho,
\ee
where $\rho = (t_1)^{M_S}$,
we can simply define the operators
\be
\hat t_i \cdot (f\; \rho) := (\hat t_i\cdot f)\;\rho      \label{t_hat_ads}
\ee
for $f \in S_q^{D-1}$, using the definition (\ref{t_hat}).
The positive definite inner product on $AdS_q^{D-1}$ is defined as 
\be
(f\; \rho,g\; \rho) = (f,g)
\label{inner_AdS}
\ee
where $(f,g)$ is given by (\ref{inner_S}).
Using (\ref{t_hat_adjoint_S}), the  adjoint of $\hat t_i$ is 
\be
(\hat t_i)^* \cdot (f\; \rho) = (\Om \hat t_j C^j_i \Om\cdot f)\; \rho;
\ee
However, we want to express this directly in terms of the Hilbert space
$AdS_q^{D-1}$, without explicitly using the above decomposition.
The result is 

\begin{theorem} (``quantum Anti--de Sitter space'')
The positive definite inner product (\ref{inner_AdS}) 
on $AdS_{q}^{D-1}$  satisfies
\berr
(u \cdot f,g) &=& (f,u^*\cdot g)                   \\
(\hat t_i f,g)     &=& (f,\hat t_i^* g)         
\err
where $u^*$ is the star (\ref{ads_involution}) on 
$\U^{res}(so(2,D-2))$, and 
\be
(\hat t_i)^* = - \Om (-1)^E \; \hat t_j C^j_i \Om (-1)^E
         = - (-1)^{E_j} \sqrt{v^{-1}} \;\hat t_j (L^-)^j_k\sqrt{v}\hat g^{ki},
\label{t_hat_adjoint}
\ee
where $E$ is the energy operator defined in Section (\ref{subsec:star_U}).
\end{theorem}
\begin{proof}
Using (\ref{R_formula}), one can see that for 
$\rho \in L^{fin}(\sum z_i M_i \L_i)$, 
\be
(\RR_1 \cdot f)\tens (\RR_2 \cdot \rho) = q^{(\l_f,\l_z)} f \tens \rho
\label{ev_R_rho_1}
\ee
where $\l_f$ is the weight of $f$ and $\l_z= \sum z_i M_i \L_i$; this 
is because the term 
$\sum \U^{+res} \tens \U^{-res}$ in (\ref{R_formula}) does not contribute
since $X_i^{\pm} \cdot \rho = 0$, and the ``large'' generators vanish as well,
by the more explicit formulas in \cite{tolstoi,kirill_resh}.
On $AdS_q^{D-1}$, one has $q^{d_i M_i z_i} = s_i$ 
(no sum) by with $s_i$ as in (\ref{ads_involution}). Hence 
$q^{(\l_f,\l_z)} f = (-1)^{E}\cdot f$, where $E$ is the
energy operator as defined in Section \ref{subsec:star_U}. Therefore
\be
(\RR_1 \cdot f)\tens (\RR_2 \cdot \rho) =  ((-1)^{E}\cdot f) \tens \rho = 
  (\RR_2 \cdot f)\tens (\RR_1 \cdot \rho)
\label{ev_R_rho}
\ee
Now 
\berr
\Om (-1)^E \hat t_j C^j_i \Om (-1)^E \cdot (f \rho) &=& 
         \Om (-1)^E \hat t_j C^j_i \sqrt{v^{-1}}  
     (\RR^{-1}_1 \tw (-1)^E \cdot f) (\RR^{-1}_2 \tw (-1)^E \cdot \rho) \nn\\
      &=&  \Om (-1)^E \hat t_j C^j_i \sqrt{v^{-1}}
                (\tw \cdot f) (\tw (-1)^E \cdot \rho) \nn\\
      &=& - (\sqrt{v} \RR_2 \tw^{-1} (-1)^E \hat t_j C^j_i \sqrt{v^{-1}}
                                                   \tw \cdot f) 
           (\RR_1 \tw^{-1} (-1)^E \tw (-1)^E \cdot \rho) \nn\\
      &=& - (\sqrt{v} \tw^{-1} \hat t_j C^j_i \sqrt{v^{-1}}
                                             \tw \cdot f) \rho\nn\\
      &=& - (\Om  \hat t_j C^j_i \Om  \cdot f) \rho.
\err
Here we used $\Om = \sqrt{v^{-1}} \tw = \sqrt{v} \tw^{-1}$, 
$\Delta(\tw^{-1}) = \RR_{21} \tw^{-1} \tens \tw^{-1}$,
and the fact that 
\be
\sqrt{v}\; \hat t_i \sqrt{v^{-1}} (f \tens \rho) = 
 - (\sqrt{v}\; \hat t_i \sqrt{v^{-1}} f) \tens \rho
\ee
for 
$f \rho \in AdS_q^{D-1}$, which is not hard to see.
\end{proof}

This completes the definition of the star--algebra of coordinate functions on
the real quantum Anti--de Sitter space $AdS_q^{D-1}$, extended by $\Om$ or
$\U^{res}$. In the second form of (\ref{t_hat_adjoint}), 
the classical limit $q \rightarrow 1$ is easy to understand
for low--energy states (i.e. states with weight close to the origin),
since then $(L^-)^j_k$ becomes $\d^j_k$, and the star approaches
the classical form 
$\hat t_i^* = -\hat t_j (-1)^{E_j} g^{ji}$ in the limit, see (\ref{ads_real}).
For energies $E < M_S$, the scalar fields have the same structure as 
classically. Self--adjoint position operators can be defined as
$\hat t_i + \hat t_i^*$, and together with $\U$ we have a complete framework
for quantum mechanics on the $q$--deformed $AdS$ space. 

It should be noticed that  the operators $\hat t_i$
satisfy the algebra (\ref{euclid_CR}) on all $\F(n)$ except the one
with maximal $n$. This is related to the classical issue
of the domain of definition of the position operators on AdS space,
which are then
unbounded operators, and can be obtained as the limit 
of $\hat t_i$ for $q \rightarrow 1$. Then the ``maximal''
$\F(n)$ becomes the lowest--weight module with minimal
mass, and could be called the ``massless'' field
(even though the definition of ``massless'' is somewhat ambiguous on
the AdS space). In the classical case, the product of this field 
with the coordinate function $t_i$ 
is not in the Hilbert space any more, and one might simply say that 
this field is not in the domain of definition of the position operator.  
However its domain can be extended, and 
the definition (\ref{t_hat}) of the operator $\hat t_i$ amounts to 
an extension such that (\ref{t_hat_adjoint}) still holds, at the expense
of the commutation relations. Hence we enter the nontrivial subject of 
extensions of the domain of definition of symmetric
operators. This shows that even issues of operator analysis can be addressed
in our ``regularized'', finite--dimensional version, 
which might be of interest even from the purely classical point of view.

\sect{Further developments}

\subsection{Calculus}
\label{subsec:calculus}

It is well--known that there exists a $q$--analogue of the 
usual differential calculus, differential forms etc. on the 
quantum Euclidean space; for an extensive review see e.g. \cite{fiore}.
Even though this is 
usually done for generic $q$, there is no problem extending it to 
roots of unity. Moreover, the reality structures defined in 
Section 5 immediately generalize to these algebras. The basic relations
are as follows \cite{WZ}:
\be
\dl_i x_j = {g}_{ij} + q_S (\hat R^{-1})_{ij}^{kl} x_k \dl_l,
\label{dl_x_cr}
\ee
with
\be
(P^-)^{ij}_{kl} \dl_i \dl_j =0.
\ee
The one--forms $dx_i$ satisfy
\be
dx_i dx_j = -q_S \hat R_{ij}^{kl} dx_k dx_l, \quad x_i dx_j = 
   q_S \hat  R_{ij}^{kl} dx_k x_l.
\ee
All this is covariant under the symmetry algebra $\U^{res}$. 

The differential forms that should be used on the sphere
or Anti--de Sitter space are $\frac 1r dx_i$, where $r^2 = x^2$. 
However, the number of independent one--forms remains $D$ 
instead of $D-1$, since $d(x^2)$ cannot be omitted (i.e. factored out) 
consistently. This may in fact be quite
interesting from a physical point of view, see Section
\ref{subsec:physic_mod}.
From a formal point of view, this is related to the fact that 
there exists a certain analog of the Dirac--operator in the sense of 
Connes, 
\be
w = \frac {q_S^2}{(q_S+1)r^2} d(r^2) 
\ee
which generates the calculus on these quantum spaces by 
\be
[w, f]_{\pm} = (1-q_S) df
\ee
for any form $f$ with the appropriate grading. This was discovered 
by Bruno Zumino \cite{zumino_priv}. It satisfies 
\be
dw = w^2 = 0.
\ee
It is also worth observing that $d(x_1)^{M_S} =0$.
One could go on and define an analog of the Laplace operator by
$\Delta = g^{ij} \dl_i \dl_j$, and 
write down wave equations. However 
at roots of unity, this is perhaps a bit too simple--minded.
A different proposal will be made in Section \ref{subsec:physic_mod}.


\subsection{Integral}
\label{subsec:integration}

The invariant integral of polynomial functions on the 
quantum sphere 
is uniquely given by the projection on the invariant component 
(\ref{invar_state}), as in \cite{integral,weich,fiore} for real $q$. 
To make the notation more intuitive, we can write
\be
\int_{S_{q}^{D-1}} f(\hat t) = <f(\hat t)>
\ee
where $S_q^{D-1}$ is considered as the algebra
generated by $\hat t_i$. One can give an explicit formula,
using the calculus discussed in the previous section:
Consider the invariant linear functional \cite{integral,fiore}
defined on $M_q^{2k}$ by
$<f(x)>_{\Delta}:= c_k^{-1} \Delta^k f(x)_{\bb}$  for 
$k < k_S$, where $c_k$ is given by (\ref{c_n}) in the Appendix, 
and the vertical bar means ``evaluation'', i.e. it
anihilates the derivatives. We define $c_k$  such that 
$<x^2 g(x)>_{\Delta} = <g(x)>_{\Delta}$, hence $<\;>_{\Delta}$ is
a functional on $S_{q}^{D-1}$.
Since it is also invariant, it agrees with the unique integral, 
\be
\int_{S_{q}^{D-1}} f(\hat t) = c_k^{-1} \Delta^k f(x)_{\bb}
\label{integral_laplacian}
\ee
if $f(\hat t)$ is given by a reduced polynomial of degree $2k$; of course, 
$\int_{S_{q}^{D-1}} f(\hat t) = 0$ if $f(\hat t)$ is odd. 
(\ref{state_bar}) implies that  
\be
\Big(\int_{S_q^{D-1}} f(\hat t)\Big)^* = \int_{S_q^{D-1}} f(\hat t)^*,
\ee
with the involution (\ref{t_hat_adjoint_S}); note that even though
$f(\hat t)^*$ is an element of $\U^{res}\smash S_{q}^{D-1}$, the 
projection on the invariant component of $S_q^{D-1}$ unambiguously reduces
$\int_{S_q^{D-1}} f(\hat t)^*$ to a complex number. 
In particular, 
\be
\int_{S_q^{D-1}} f(\hat t)^* g(\hat t) 
\ee
agrees with the positive--definite inner product on $S_q^{D-1}$ defined in 
Theorem \ref{invol_thm}.

On $AdS_q^{D-1}$, the situation is similar. The algebra of $S_{q,\C}^{D-1}$
implies for example
\berr
(t_1)^{M_S} t_i &=& -t_i (t_1)^{M_S}, \quad i \neq 1,D \nn\\
(t_1)^{M_S} t_D &=& t_D (t_1)^{M_S}
\err
(the minus could be eliminated by defining $\z_i = q_S^E t_i$).
Similarly, one obtains commutation relations between the other elements of
$\F(M_S)$ and the generators $t_i$, and hence between the 
elements of $\F(M_S)$ themselves. This algebra is consistent with the
action of $\U^{res}$, which is that of $\tilde \mg$ as we have seen earlier.
Moreover, it is naturally a Hilbert space as we have used many times,
hence the algebra generated by $\F(M_S)$ is naturally a 
star algebra, using the adjoint.
The inner product can again be obtained from a unique invariant state.
Combining this with $S_q^{D-1}$, 
$\A:=(\oplus_k \F(k M_S)) \tens S_q^{D-1}$ becomes a star--algebra.
The unique invariant state defines an integral on $\A$, 
so that the inner product (\ref{inner_F}) of this Hilbert space  
can be written in the form
\be
(f,g) = \int_{\A} f^* g.
\ee
In particular, this defines a positive--definite integral on $AdS_q^{D-1}$.

\subsection{Towards formulating physical models}
\label{subsec:physic_mod}

We have already discussed the one--particle states for scalar
particles in detail: they are just the irreducible \reps 
$\F(n) \cong L^{fin}(n \l_1) \in AdS_q^{D-1}$ for $M_S \leq n < M_S+k_S$,
or equivalently  the lowest weight modules $L_{fin}(k \l_1)$ 
for $(D-2)/2 < k \leq M_S$. The integer
$k$ characterizes the ``mass'' of the particle, in the Anti--de Sitter sense.
How it should be determined is a dynamical question; one might e.g.
write down Lagrangians and try to formulate a variational principle.
In particular, this would allow to consider interaction terms,
since functions on quantum Anti--de Sitter space can be multiplied.
We will come back to this in a moment.

Spin one particles or vector fields are irreducible \reps
$L^{fin}(k \l_1 - \a_1)$, where $k$ satisfies certain unitarity bounds.
They can be described as one--forms on $AdS_q^{D-1}$, 
as discussed in Section \ref{subsec:calculus}. As we pointed out, there
are now $D$ components instead of $D-1$, the additional one corresponding
to ``radial'' forms $f(t) dr$. In the classical limit, they simply become
scalar fields on $AdS$ space, which are intimately connected to their
vector partners. In fact in the massless case, they turn out to 
be ghosts of the gauge fields. Moreover, there 
is a natural BRST operator which arises from $\U^{res}$ \cite{brst,thesis}. 
All these surprising
features show that the remarkable structures that arise at roots of unity
may have very interesting physical interpretations. 
We should also mention the classical symmetry $\tilde \mg$
here, which appears to be spontaneously broken in a 
natural way to $su(2)^r$ for $D=2r+1$, and to $u(1)^r$ for $D=2r$ 
\cite{nc_reps}. Of course, fields with different spin can also be described.

However, the correct formulation of a dynamical principle on $AdS_q$ is not
clear, and deserves some discussion. The traditional approach would 
be to write down a Lagrangian, which determines a Hamiltonian, which in
turn dictates the dynamics of a model. On $AdS_q^{D-1}$ however, the energy 
is given a priori; it is simply the element $E$ of the Cartan subalgebra
of $\U(so(2,D-2))$ defined in Section 3.1. Hence it seems that 
the dynamics should be formulated here by a principle which determines
the physical, ``on--shell'' degrees of freedom, as in string theory.
This is quite trivial
in a non--interacting model, since it just amounts to picking
an irreducible representation. In an interacting model, determining
a consistent physical Hilbert space in terms of the \reps we have 
at our disposal seems very nontrivial. In particular, it involves 
tensor products of the one--particle states, which is a rather complicated
and rich topic for quantum groups at roots of unity
\cite{CH_P}. This is also related to knot theory via the 
$\hat R$ --matrix and the Drinfeld Casimir $v$ (\ref{v}), which has proved 
to be very important in earlier sections. In view of
this and other considerations, we propose the following
``on--shell'' condition:
\be
v=1                    \label{v_1}
\ee
or perhaps $v=c$ for a constant $c$,
as an operator identity on the physical many--particle Hilbert 
space, supplemented by other conditions.
This is somewhat reminiscent of the on--shell condition in string theory, 
$L_0 =a$ (among others),
see also \cite{gomez}. It will be discussed further in the next section.

Nevertheless, there are obstacles if one tries to
perform a ``second quantization''. In particular, it is neither clear
how to impose a symmetrization postulate
on $AdS_q^{D-1}$, nor how to define an invariant path--integral.
Consider many--particle states, which should be
described by tensor products of the one--particle irreducible \reps
$L^{fin}(k \l_1)$. A first complication is that at roots of unity,
such tensor products are generally not completely reducible. 
However, there exists a reduced tensor product, which
yields precisely the correct ``physical'' many--particle states, as a direct
sum of unitary irreducible representations. This was defined in the 
case of $\U^{fin}(so(2,3))$ in \cite{ads_paper}, and will be discussed more
generally elsewhere\footnote{it is different from the usual ``truncated''
tensor product \cite{keller,CH_P,mack_schom} defined in the context of 
conformal field theory.}. The more difficult problem is that there seems to 
be no natural definition of a totally symmetric or antisymmetric 
tensor product for more than 2 factors, because the $\hat R$ --matrix
represents (a quotient of) the braid group rather than the symmetric group. 
A related difficulty arises in the context of path integrals: 
if one expands the fields as 
$\psi(t) = \sum a_{(\underline{k})} f^{(\underline{k})}(t)$ in the usual
way, then the coefficients $a_{(\underline{k})}$ should be covariant under 
$\U^{fin}$ as well, hence they cannot be ordinary numbers, and there is
no obvious definition of an integral. Some new ideas are needed here,
presumably related to the natural concepts of braiding, links and vertices 
associated to the fields $L^{fin}(k \l_1)$. A proposal 
in this direction has been given recently in a somewhat different 
context \cite{oeckl}, which is 
however not consistent with a Hilbert space formulation.

Finally, we want to mention that it is easily possible to describe fields
on covering spaces of $AdS_q^{D-1}$. This is done by giving up the 
quantization of the eigenvalues of the Energy $E$ to (half)integers,
which arises from the compact time--like curves on Anti--de Sitter space.
The corresponding unitary \reps of $\U^{fin}(so(2,D-2))$ with 
non--integral weights indeed exist \cite{nc_reps}. However, $\U^{res}$ is then
not admissible any more, which means that the classical symmetry $\tilde \mg$
disappears. The fields cannot be described simply by
polynomials in the $t_i$ any more, but they are still finite--dimensional
with similar properties.

\sect{Discussion of physical aspects}

Some remarks on the physical aspects of the suggested framework are in order.

\paragraph{Hilbert space}
First and foremost, the Hilbert spaces of particles respectively fields
on $AdS_q^{D-1}$ are finite--dimensional, even though they are 
covariant under the noncompact symmetry algebra $\U(so(2,D-2))$.
Since all local aspects of such a theory should have a smooth limit to the
flat, undeformed case, this can nevertheless
be viewed as a regularization of the classical, flat case
if one takes the limit  $R \rightarrow \infty$. 
All states have manifestly positive energy,
with a high--energy cutoff of the order of $M$, in units where $R=1$.
Moreover, there exists only a finite number of inequivalent 
one--particle representations.

The total number of states in $AdS_q^{D-1}$ as defined in (\ref{real_AdS})
can be calculated easily. For simplicity we calculate 
\berr
\dim(AdS_q^{D-1}) &=& \dim(S_q^{D-1}) \approx \dim(\sum_{0 \leq k < M_S} \F(k))
          = \dim(M_q^{M_S-1}) + \dim(M_q^{M_S-2}) \nn\\
  &=& \(\begin{array}{c} M_S+D-2 \\ D-1 \end{array}\) + 
      \(\begin{array}{c} M_S+D-3 \\ D-1 \end{array}\)  
  \approx 2\frac{M_S^{D-1}}{(D-1)!},
\err
for large $M_S$.  This means that if we assign the classical area to
the sphere $S_q^{D-1}$, then the number of states per area is
approximately
\be
\frac{\# states}{area} \approx c \big(\frac{M_S}{R}\big)^{D-1},
\ee
where $c \approx \sqrt{2} \big(\frac{e\pi}{2D}\big)^{D/2}$.
This corresponds to a ``quantum length'' of $L_{min} = R/M_S$,
and the same should hold on $AdS_q^{D-1}$.

However, 
the physical interpretation of the cutoff $M_S$ is not entirely clear.
At first sight, on may be tempted to relate $L_{min}$ to the Planck scale,
even though we have not attempted to formulate gravity here.
On the other hand, the commutation relations (\ref{euclid_CR}) together
with (\ref{t_2}) are inhomogeneous and
suggest the existence of another, larger length scale
given by
$L_{NC}^2 \approx R^2 i(q-q^{-1})$ or $L_{NC}/R \approx \sqrt{M^{-1}}$,
where one would expect the noncommutativity to become important. 
This implies a certain 
quantization of the radius $R$ of AdS space in terms of $L_{NC}$ and $M$.
Of course this argument may be questioned because the coordinate
functions $t_i$ are not real. However this scale is relevant
whenever functions on $AdS_q^{D-1}$ are multiplied, for example if one
would write down
interaction terms in Lagrangians, as discussed in the previous section.
We will see further evidence for the importance of this scale in the 
next paragraph. Hence $L_{NC}$ is expected to be the relevant 
``non--commutative'' scale where the ``fuzzyness'' of the space 
becomes important, while 
$M_S/R$ is the maximal energy of the entire many--particle Hilbert space.
Of course, these issues can only be settled once a physical model has been
formulated and studied. It is nevertheless quite interesting 
that in the context of
loop quantum gravity and spin networks \cite{baez}, a similar quantization
of the cosmological constant in terms of the Planck length and the
root of unity of a quantum group was found.

At this point, we can get some insight into the proposed on--shell condition 
$v=1$ (\ref{v_1}), for the one--particle sector. Consider the 
unitary \reps $L^{fin}(\l) = L_{fin}(\mu)$ for $\l = 2M_S \l_1 - \mu$, 
which has 
lowest weight $\mu$. It satisfies $v=1$ if the classical Casimir 
$c_{\l}$ is divisible by $2 M$, by (\ref{v_h_w}). 
This holds in particular if $c_{\l} = 0$, which is a good
definition of massless fields on $AdS$ space; for example, it means
$\mu = (D-2)\l_1$ for scalar fields. It also characterizes certain
gauge fields with higher spin \cite{ferrara},
for example the massless vector fields on $AdS_q^4$, 
which contain the correct degrees of freedom of gauge fields \cite{ads_paper}.
This will be discussed in more detail elsewhere; in any case,
the constraint $v=1$ yields certain massless fields, and a series of
massive fields with masses of the order of $\sqrt{M}$, since $c_{\l}$ is 
quadratic in $\l$. The latter is 
essentially $L_{NC}^{-1}$, the scale found above 
where noncommutative effects are expected to
become important. Therefore one gets qualitatively a similar picture 
as in string theory. However, the tower of massive fields terminates here.

\paragraph{Anti--particles}
Another interesting issue is the physical meaning of certain observables, 
such as the
operator $E$ defined in Section 3.1. For $E$ much smaller than $M_S$, 
this is clearly the generalization of the classical energy. However, note that
the spectrum (=the weights) of the irreducible \reps 
$L^{fin}((M_S+k) \l_1)$ for 
$k < M_S$ is symmetric around $M_S \l_1$, and there are as many 
states with $E=2M_S-n$ as there are with $E= n$. Hence one could
interpret $[E]_{q_S} \approx \frac {M_S}{\pi} \sin(\frac{E\pi}{M_S})$ 
as the physical 
energy which has a maximum at $E=M_S$, and interpret the states with 
$E > M_S$ as anti--particle states. This is actually
more natural from the mathematical point of view. Then particle and 
anti--particle states could be unified in one irreducible representation,
which is a fascinating picture. Alternatively, one may interpret 
$L^{fin}((-M_S+k) \l_1) \subset L^{res}((M_S+k) \l_1)$ as the anti-particle 
states corresponding to $L^{fin}((M_S+k) \l_1)$. 
Then the operator $\Om$ defined in (\ref{Omega}) plays the 
role of a $CPT$ operator. It could also be redefined
using $U(\tilde \mg)$ so that it
leaves $L^{fin}((M_S+k) \l_1)$ invariant, if latter is interpreted 
as containing both particle and antiparticle states as just discussed.

\paragraph{Comments on the AdS--CFT correspondence}
Recently, an interesting conjecture between string or M theory 
on $AdS^{n} \times W$ and (super)conformal field theories in $n-1$ 
dimensions has been proposed \cite{maldacena}. 
This includes the cases
$AdS^{5} \times S^{5}$, $AdS^{3} \times S^{3} \times M$, 
and $AdS^4 \times S^7$.
Moreover, this has been extended to certain quotients (orbifolds)
$S^n / \Gamma$ \cite{kachru}, which can break supersymmetry. 
Even though no precise definition of the string theory or M theory
side is known, it has been argued that it implies a 
``stringy exclusion principle'' \cite{strominger}, which may indicate
that  the ``effective'' space--time is noncommutative. 
One may try to identify the integer $M$ with the number of $D$--branes, 
as was already suggested in the literature \cite{ho,ramgoolam}.
Hence it is quite striking that the mere definition of quantized
AdS spaces at roots of unity presented here leads naturally to such 
product spaces with spheres, 
as explained in Section \ref{subsec:product_spaces}.
In particular, the appearance of $AdS^4 \times S_{\chi}^7$ was completely 
unexpected, and occurs without making any choices.
Another feature of the AdS--CFT correspondence is the 
quantization of the radius of $AdS$ space
in terms of the Planck scale and the number of $D$--branes \cite{maldacena}.
We found a similar quantization above, where the relevant integer is $M$.

One may object that there is no reason why precisely
the Drinfeld--Jimbo quantum groups at roots of unity should play a 
role in the AdS--CFT correspondence. On the other hand, it has been
known for a long time \cite{gomez} that they do play a role in certain
conformal field theories, such as the WZW models. 
Since the AdS space can be viewed as a coset of the AdS group,
it is plausible that the quantum AdS group at roots of unity
should be important in the above context. Hence it seems worthwhile
to study the consequences of that assumption; as we pointed out, they
are quite striking and rich. 
However at present, it is difficult to study these issues in more detail,
since little is known about both the string theory side and the 
quantum group side.

\section{Acknowledgements}

I would like to thank P. Aschieri, H. Grosse, J. Madore, J. Pawelczyk, 
G. Reiter, E. Scheidegger, S. Theisen and Z. Yin 
for interesting  discussions, and J. Wess for trustful support. 
I am also grateful to C.--S. Chu, P.--M. Ho, B. Morariu and B. Zumino for 
discussions in the early stages of this work.

\sect{Appendix A:Vector \reps}

\begin{figure}
 \epsfxsize=3in
   \hspace{1.7in}
   \epsfbox{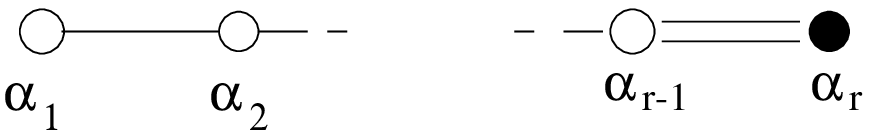}
   \caption{$so(2r+1)$}
\label{fig:B_r}
\end{figure}

We review the vector \reps for convenience. First consider $so(D)$ with 
$D=2r+1$, and Dynkin diagram as in figure \ref{fig:B_r}.
Then $V_D = L^{res}(\L_1)$ has a basis $x_i$ for $i=1,...,D$ with weights
$\l_i = \sum_j (\l_i,\ad_j) \L_j$. The coordinates $(\l_i,\ad_j) \in \Z$ 
will be listed below as row vectors. For $D>3$, the weights are 
in descending order
\be
\begin{array}{cll}
\l_1 &= \L_1        &= (1,0,0,...,0) \\
\l_2 &= \L_1 - \a_1 &= (-1,1,0,...,0)  \\
\l_3 &= \L_1 - \a_1-\a_2 &= (0,-1,1,0,...,0) \\
... & & \\
\l_{r} &= \L_1 - \sum_{i=1}^{r-1} \a_i &= (0,...,-1,2) \\
\l_{r+1} &= \L_1 - \sum_{i=1}^{r} \a_i &= (0,...,0) \\
\l_{r+2} &= \L_1 - \sum_{i=1}^{r} \a_i -\a_r &= (0,...,1,-2) \\
\l_{r+3} &= \L_1 - \sum_{i=1}^{r} \a_i -\a_r -\a_{r-1} &= (0,...,1,-1,0)\\
... & & \\
\l_{2r+1} &= \L_1 - 2\sum_{i=1}^r \a_i &= (-1,0,...,0) = -\L_1. 
\end{array}
\ee
If $D=3$, then $\l_1 = 2 \L_1 = \a$. In any case, 
$\l_1 = \sum_{i=1}^r \a_i$, with $(\l_1,\l_1) = 2$. 
$\a_r$ is the short root, and\footnote{in \cite{FRT}, the conventions 
are such that there is an additional factor $1/2$.}
$\rho_i = (\rho,\l_i) = (2r-1,2r-3,...,1,0,-1,..,-2r+1)$. 

\begin{figure}
 \epsfxsize=3in
   \hspace{1.7in}
   \epsfbox{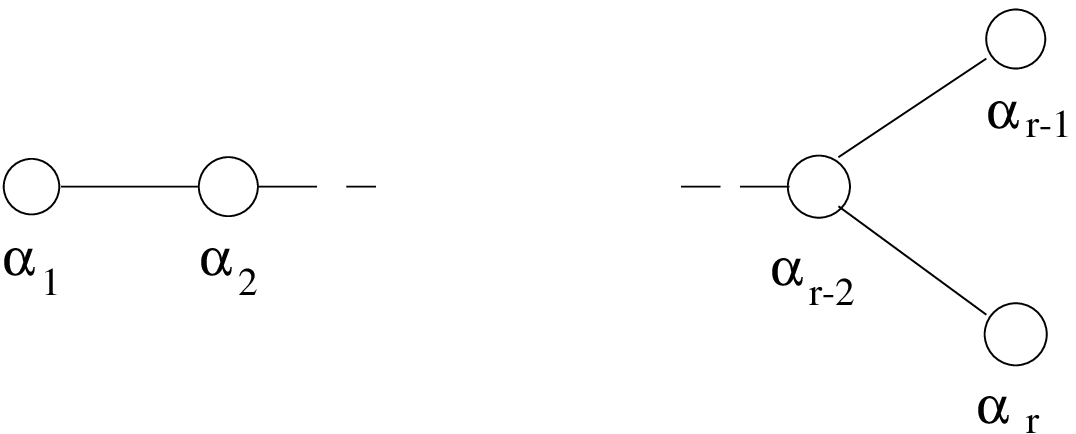}
   \caption{$so(2r)$}
\label{fig:D_r}
\end{figure}

Next consider $so(D)$ with $D=2r$ and Dynkin diagram as in 
figure \ref{fig:D_r}.
The vector \rep $V_D$ has a basis $x_i$ for $i=1,...,D$ with weights
$\l_i$. For $D>4$,  the weights in descending order are now
\be
\begin{array}{cll}
\l_1 &= \L_1        &= (1,0,0,...,0) \\
\l_2 &= \L_1 - \a_1 &= (-1,1,0,...,0)  \\
\l_3 &= \L_1 - \a_1-\a_2 &= (0,-1,1,0,...,0) \\
... & & \\
\l_{r-1} &= \L_1 - \sum_{i=1}^{r-2} \a_i &= (0,...,-1,1,1) \\
\l_{r} &= \L_1 - \sum_{i=1}^{r-2} \a_i -\a_{r-1} &= (0,...,-1,1) \\
\l_{r+1} &= \L_1 - \sum_{i=1}^{r-2} \a_i -\a_r &= (0,...,1,-1) \\
\l_{r+2} &= \L_1 - \sum_{i=1}^{r} \a_i   &= (0,...,1,-1,-1) \\
\l_{r+3} &= \L_1 - \sum_{i=1}^{r} \a_i -\a_{r-2} &= (0,...,1,-1,0,0)\\
... & & \\
\l_{2r} &= \L_1 - \sum_{i=1}^{r} \a_i -\sum_{i=1}^{r-2} \a_i &= (-1,0,...,0) 
         = -\L_1. 
\end{array}
\ee
If $D=4$, then $\l_1 = \L_1 + \L_2 = \frac 12 (\a_1 + \a_2)$. 
In any case, $\L_1 = \sum_{i=1}^r \a_i - \frac 12 (\a_{r-1}+ \a_r)$,
and $(\l_1,\l_1) = 1$. The Weyl vector is invariant under
the automorphism $\g$, and 
$\rho_i = (\rho,\l_i) = (r-1,r-2, ...,1,0,0,-1,...,-r+1)$.

\sect{Appendix B: On the Weyl Element $\w$}

We want to give some explanations to the remarkable
formulas (\ref{del_om}), (\ref{w_2}) and (\ref{autom_w}). 

To get some confidence in these formulas, notice that if 
\be
\Del(\w) = \RR^{-1} \w\tens \w
\label{del_om_left}
\ee
holds on the tensor product of 2 fundamental representations, then 
it holds for
any representations. This can be seen inductively: if (\ref{del_om_left})
holds on $V\tens V'$, then it also holds on the \reps in 
$V\tens V'\tens V''$. Indeed by coassociativity, 
\be
({\id}\tens \Del)\Del(\w) \stackrel{!}{=} \RR^{-1}_{1,(23)} (\w\tens \w)
\ee 
acting on $V\tens(V'\tens V'')$ should be the same as
\be
(\Del\tens {\id})\Del(\w) \stackrel{!}{=} \RR^{-1}_{(12),3} (\w\tens \w)
\ee
acting on $(V \tens V')\tens V''$. The rhs agree indeed since $\U$ is
quasitriangular, see e.g. Lemma 1.1.1 in \cite{thesis};
thus the lhs agree as they should.
Furthermore, since the action of $\w$ on any given finite--dimensional
\rep can be expressed in terms of generators of $\U$, it satisfies 
$\Del'(\w) = \RR \Del(\w) \RR^{-1}$, and together with (\ref{del_om_left})
this implies $\Del(\w) = \w\tens \w \RR_{21}$. (\ref{del_om_left})
also shows that $\w$ is well-defined at roots of unity; this
can alternatively be seen  from the explicit formulas 
given e.g. in \cite{lev_soib}. 

The action of $\w$ on the fundamental \reps can be found 
explicitly, see \cite{kirill_resh}. If they are real, it is essentially 
the invariant metric $g_{ij}$, by (\ref{g_JC}).

(\ref{w_2}) together with the statement that $\epsilon = \pm 1$ on any 
irreducible \rep
now follows easily, since $\epsilon := v^{-1} \w^2$ is grouplike,
i.e. $\Del(\epsilon) = \epsilon\tens\epsilon$, and it only remains
to check that $\epsilon$ is $\pm 1$ on the fundamental representations, where
it agrees with the classical limit. 

Finally, we show $\w x \w^{-1} = \theta S \g(x)$. Again, this
can be verified for the fundamental representations, and it follows for
tensor product \reps by (\ref{del_om_left}) and the fact that both
sides of (\ref{autom_w}) are algebra maps. 

\sect{Appendix C: some proofs}

\paragraph{Proof of Theorem  \ref{scalar_thm}}

The fact that $\F(k) \cong L^{res}(k \l_1) = L^{fin}(k \l_1)$ 
have the same 
character as classically for $k$ as stated was already explained in the text; 
this can also be seen to a large extent by considering various Casimirs. 
It remains to show that they are linearly independent at 
roots of unity, and that (\ref{M_k_decomp}) holds.
Linear independence is clear: if there were a nontrivial linear combination
of vectors in the $\F(k)$ which is zero, then there would exists a nontrivial 
submodule of the form
$\F(k_0) \cap (\oplus \F(k_i)) \neq 0$. This would mean that at least one
$F(k)$ contains a highest weight vector, which is in contradiction
to $\F(k)$ being irreducible (this also follows from well--known facts
about \reps at roots of unity \cite{CH_P}). Now (\ref{M_k_decomp})
follows immediately, because the characters 
and hence the dimensions of the $\F(k)$ are the same as classically.

\paragraph{Proof of Lemma \ref{orth_lemma}:}

Assume that $\g$ acts trivially on $\l$. Since $L^{res}(\l)$ is a 
highest weight module, every $v \in L^{res}(\l)$ can be written in the
form $v = u^- \cdot w_{\l}$ for $u^- \in \U^{- res}$. Define 
$T_{\g} \cdot v := \g(u^-) \cdot w_{\l}$. This is well--defined, because 
$T_{\g}\cdot (H_i\cdot w_{\l}) = \g(H_i)\cdot w_{\l} = H_i \cdot w_{\l}$
by the assumption. Clearly $T_{\g}^2 =1$, and if $C$ is the matrix 
representing $T_{\g}$, then $C^2 = 1$ and (\ref{C_conj}) follow. 

To see that  $L^{res}(\l)$ is real, observe that
$L^{res}(\l)$ contains a vector $\om \cdot w_{\l}$ with weight $-\l$, 
since $\w H_i \w^{-1} = - \g(H_i)$ and $\g$ acts trivially on $\l$.
This implies (e.g. by the method of characters) that
for generic $q$, the \rep $L^{res}(\l) \tens L^{res}(\l)$ 
has precisely one trivial component, which defines the invariant tensor. 
It can be viewed as a solution of the equations $X_i^+ = H_i = 0$
which are analytic in $q$. Therefore this invariant tensor is smooth, 
and induces an invariant tensor also at roots of unity, where the irreducible 
\rep $L^{res}(\l)$ can be obtained as quotient of the generic ones. 
Hence $L^{res}(\l)$ is real. This can also be seen from Theorem 
\ref{L_res_theorem}.

Next we prove (\ref{orthogonal_basis}).
(\ref{A_eqn}) implies  
\be
\pi(u)^T = B^{-1} \pi(\theta(u))B         \label{B_eqn}
\ee
with $B = CA$, and applying this twice yields
\be
\pi(u) = (B^T B^{-1}) \pi(u) (B^T B^{-1})^{-1},
\ee
using $\theta^2 = \id$. 
Since the \rep is irreducible, Schur's Lemma implies that
$B^T B^{-1} = \a \one$, or equivalently $B^T = \a B$ for some $\a \in \C$.
This is possible only for $\a = \pm 1$, thus
\be
B^T = \pm B.
\label{B_T}
\ee
Similarly, taking the complex conjugate of (\ref{B_eqn}), we obtain
\berr
\pi(u)^{\dagger}  &=& {B^{-1}}^* \pi(\theta(u))^* B^*
                                           \nonumber\\ 
           &=& {B^{-1}}^* \pi(\theta(u^*))^T B^*    
    =  {B^{-1}}^* B^{-1} \pi(u^*) B B^*,
\err
using (\ref{unitary_basis1}) and $\theta(u)^* = \theta(u^*)$.
On the other hand, $\pi(u)^{\dagger} = \pi(u^*)$, and it follows
as above that $B B^* = \b \one$. Together with (\ref{B_T}),
this implies that $B B^{\dagger} = \pm \b \one$. 
The constant $\pm \b$ must be real and positive because $B B^{\dagger}$
is a positive definite matrix. We can assume that the determinant of
$B$ (or $g$) has modulus one, hence  
\be
B B^{\dagger} = \one.                      
\ee
Decomposing $B = X + i Y$ with real matrices $X,Y$, it follows that 
$[X,Y] =0$. Thus $X$ and $Y$ can be simultaneously
(block--)diagonalized with a real, orthogonal matrix $O$, 
and $B$ can be written
in the form $B = O^T J O$ with $J$ diagonal if $B^T = B$, or $J$ 
block--diagonal with blocks proportional to
$\(\begin{array}{cc} 0 & -1 \\ 1 & 0 \end{array} \)$ if $B^T = -B$.
Changing the basis using the matrix $O$ yields 
$\pi(u)^T = J^{-1} \pi(\theta(u)) J$
while preserving (\ref{unitary_basis1}),
and $J$ has the desired form after a transformation with a unitary 
diagonal matrix. In particular, (\ref{orthogonal_basis}) together with 
$J^T C = g \pi(\tw)$ (see below (\ref{A_eqn})) implies that 
\be
\hat g = J^T C \pi(\tw^{-1}).                     \label{g_JC1}
\ee

Next, we show that $J=1$ if the \rep is orthogonal at $q=1$, 
and $J^T = -J$ if it is symplectic.
Equation (\ref{g_invar}) holds also for $u=\tw$, where $\eps(\tw) = 1$
(this can be understood e.g. from the explicit form \cite{jantzen} of 
$\w$ as a power series in $\U^{res}$, or simply by the
fact that $\tw$ acts trivially on trivial representations).
Using (\ref{del_om}), this implies 
\be
g_{ij} (R^{-1})^{ij}_{mn}\pi^m_k(\tw) \pi^n_l(\tw) = g_{kl}
\ee 
Since $L^{res}(\l)$ is real,
$L^{res}(\l) \tens L^{res}(\l)$ contains precisely one trivial module 
for generic $q$, as classically. In that case, it is known \cite{resh_1}
that 
\be
g_{ij} (R^{-1})^{ij}_{mn} = g_{ij} (\hat R^{-1})^{ij}_{nm} 
        = \sigma q^{c_{\l}} g_{nm}
\label{w_inv}
\ee
where $c_{\l}$ is the classical quadratic Casimir.
Now for $q=1$, $(\hat R)^{ij}_{nm} = \d^{ij}_{mn}$ is the 
permutation operator, hence $\sigma=1$ if $L(\l)$ is orthogonal at $q=1$, and 
$\sigma=-1$ if it is symplectic.
In any case, $\sigma q^{c_{\l}} g_{nm} \pi^m_k(\tw) \pi^n_l(\tw) = g_{kl}$,
and $\sigma q^{c_{\l}}g_{nm}\pi^m_k(\tw) = g_{kl}\pi^l_n(\tw^{-1})$.
Now $\tw^{-1} = \tw v^{-1}$ by (\ref{w_2}), and together with (\ref{v_h_w})
we obtain $g_{nm}\pi^m_k(\tw) = \sigma g_{kl}\pi^l_n(\tw)$, or 
$g \pi(\tw) = \sigma (g \pi(\tw))^T$. By (\ref{g_JC1}), this implies
$J C = \sigma (JC)^T$. By definition, $C$ leaves the subspace
with weight $\pm \l$ invariant, where $C=1$. Hence $J^T = \sigma J$, 
and the claim follows.  

In particular, we have shown $C^T = J C J^{-1}$,
which can also be seen by calculating $\pi(\g(u))^T$ 
in 2 different ways and using Schur's Lemma.
Finally, taking the adjoint of (\ref{C_conj}) yields 
$\pi(\g(u))^{\dagger} = \pi(\g(u^*))= C\pi(u)^{\dagger}C$.
On the other hand, 
$\pi(\g(u))^{\dagger} = C^{\dagger}\pi(u)^{\dagger}C^{\dagger}$,
and Schur's Lemma implies that $C C^{\dagger} = \a \one$
for some $\a \in \C$. Since $C=1$ on the highest weight vector, we obtain
$C^{\dagger} = C$.   $\quad \Box$

(One can show similarly that there exists another basis such that
the \rep satisfies both (\ref{unitary_basis1}) and 
$\pi(u)^T = - J^{-1} \pi(u) J$ for $u \in \{X_i^{\pm}, H_i\}$.)

\paragraph{Proof of Proposition \ref{involution_theorem}} 

First we check the compatibility with the cross product algebra.
Applying the bar to $u x_i = u_1\cdot x_i u_2$ where 
$\Delta(u) = u_{1}\tens u_{2}$
is the Sweedler notation for the coproduct, we get
\berr
\tw x_k C^k_i \tw^{-1}\obar{u} &=&\obar{u_2} \;\obar{x_l}\pi^l_i(u_1)^*\nn\\
  &=& \obar{u_2} \tw x_k C^k_l \tw^{-1} \pi^i_l(\obar{u_1}),
\err
using the fact that $V_D$ is a unitary \rep of the compact form.
Multiplying from the left with $\tw^{-1}$ and from the right with $\tw$,
we have to show
\be
x_k C^k_i \tw^{-1} \obar{u} \tw \stackrel{!}{=}
      \tw^{-1}(\obar{u})_1 \tw x_k C^k_l \pi^i_l((\obar{u})_2),
\ee
or using (\ref{autom_tw}) 
\berr 
x_k C^k_i \theta S^{-1} \g(\obar{u}) &\stackrel{!}{=}& 
     (\theta S^{-1}\g(\obar{u})_1)_1
     \cdot x_k C^k_l (\theta S^{-1}\g(\obar{u})_1)_2 \pi^i_l((\obar{u})_2)\nn\\
   &=& (\theta S^{-1}\g(\obar{u})_{12})
      \cdot x_k C^k_l \theta S^{-1}\g(\obar{u})_{11}\pi^i_l((\obar{u})_2)\nn\\
   &=& x_n C^k_l \pi^k_n(S^{-1}\g(\obar{u})_{12}) \pi^i_l((\obar{u})_2)
                                         \theta S^{-1}\g(\obar{u})_{11} \nn\\
   &=& x_n \pi^i_m((\obar{u})_2 S^{-1} (\obar{u})_{12}) C^m_n
                                       \theta S^{-1}\g(\obar{u})_{11} \nn\\
   &=& x_n \pi^i_m(\eps((\obar{u})_2)) C^m_n \theta S^{-1}\g(\obar{u})_1 \nn\\
   &=& x_n \theta S^{-1}\g(\obar{u}) C^i_n \nn\\
   &=& x_n \theta S^{-1}\g(\obar{u}) C^n_i
\err
as desired, using Lemma \ref{orth_lemma}, $C=C^T$ and standard
properties of Hopf algebras (recall that $C=1$ for $B_n$, which
simplifies this calculation).

Next we verify that $\obar{\obar{x_i}}$ is given by the conjugation 
with $v$, using the form (\ref{invol_explic}).
Applying the bar to (\ref{invol_explic}), we find using
$\obar{{L^-}^i_j} = {L^+}^j_i$ (which follows from 
$\obar\RR_1\tens \obar\RR_2 = \RR_2^{-1} \tens \RR_1^{-1}$)
and $g^{\dagger} = g^{-1}$ 
\berr
\obar{\obar{x_i}} &=& \hat g_{ik} {L^+}^k_l x_t {L^-}^t_u \hat g^{ul} \nn\\
   &=& \hat g_{ik} x_r \pi^r_t({L^+}^k_s){L^+}^s_l {L^-}^t_u \hat g^{ul} \nn\\
   &=& \hat g_{ik} x_r R^{rk}_{ts} {L^+}^s_l {L^-}^t_u \hat g^{ul} \nn\\
   &=& \hat g_{ik} x_r {L^-}^r_t {L^+}^k_s R^{ts}_{ul} \hat g^{ul} \nn\\
   &=& \hat g_{ik} x_r {L^-}^r_t {L^+}^k_s \hat g^{st} q^{-c_{\L_1}}, \nn\\ 
\err
where we used the standard commutation relations of the $L^{\pm}$--matrices 
\cite{FRT}, and 
$R^{ts}_{ul} \hat g^{ul} = \hat R^{st}_{ul} \hat g^{ul} = 
q^{-c_{\L_1}} \hat g^{st}$.
On the other hand, using (\ref{v_coprod}) one has
\berr
v x_i v^{-1} &=& x_j \pi^j_k(\RR_1^{-1}) \pi^k_i(\RR_b^{-1})\RR_2^{-1}
                 \RR_a^{-1} q^{-c_{\L_1}} \nn\\
             &=& x_j {L^-}^j_k S{L^+}^k_i q^{-c_{\L_1}} \nn\\
             &=& x_j {L^-}^j_k \hat g_{it} {L^+}^t_l \hat g^{lk} q^{-c_{\L_1}},
\err 
since 
$(1\tens \pi^k_i) \RR^{-1} = S{L^+}^k_i = \hat g_{it} {L^+}^t_l \hat g^{lk}$,
see \cite{FRT}. Clearly both forms agree. $\quad \Box$

\paragraph{Positivity of (\ref{inner_S}):}
Consider 
\berr
((t_1)^n,(t_1)^n) &=& <\obar{(t_1)^n} \sqrt{v}\cdot (t_1)^n>     \nn\\
           &=& <(t_1)^n \tw^{-1} \sqrt{v}\cdot (t_1)^n> 
\err
where we used $C^i_1 =\d^i_1$ since $t_1$ is the highest--weight vector
of $V_D$ (we omit the hat on the $t_i$ since $n < k_S$). 
Using  $(\RR_1^{-1}\cdot t_1)(\RR_2^{-1}\cdot t_1) = q^{-(\l_1,\l_1)}(t_1)^2$,
one can show that $\tw \cdot (t_1)^n = q_S^{-n(n-1)/2}(\tw\cdot t_1)^n$, hence
\berr
((t_1)^n,(t_1)^n) &=& q_S^{n(n-1)/2}<(t_1)^n\sqrt{v}(\tw^{-1}\cdot t_1)^n>\nn\\
                  &=& q_S^{n(n-1)/2} q_S^{-(n^2+n(D-2))/2}
                                       <(t_1)^n (\tw^{-1}\cdot t_1)^n>\nn\\
                  &=& q_S^{-n(D-1)/2}  <(t_1)^n (t_kC^k_l \hat g_{l1})^n>\nn\\
                  &=& q_S^{-n(D-1)/2}  <(t_1)^n (t_D \hat g_{D1})^n> \nn\\
                  &=& (g_{D1})^n <(t_1)^n (t_D)^n>
\err
where $\hat g_{ij} = g_{ij} q_S^{(D-1)/2}$, see Proposition 
\ref{involution_theorem}.
To proceed, we use the formula 
$<(t_1)^n (t_D)^n>= c_n^{-1}\Delta^n (x_1)^n {(x_D)^n}_{\bb}$  
of Section \ref{subsec:integration}, as well as 
$\Delta x_i = q_S^2 x_i \Delta + \mu \dl_i$ where
$\mu = 1+q_S^{2-D}$. Using $\hat R^{11}_{11} = q_S$ \cite{FRT},  
(\ref{dl_x_cr}) implies $\dl_1 x_1 = x_1 \dl_1$, and therefore
\be
\Delta x_1^n = q_S^{2n} x_1^n \Delta + q_S^{n-1} [n]_{q_S}\mu x_1^{n-1}\dl_1. 
\ee
Now observe that $\dl_i {x_D^n}_{\bb} =0$ unless $i=1$, hence
$\dl_1 {x_D^n}_{\bb} = (g_{1D} + q_S^2 x_D \dl_1) {x_D^{n-1}}_{\bb}$, 
using $(\hat R^{-1})^{D1}_{1D} = q_S$. Therefore
\be
\dl_1 {x_D^n}_{\bb} = q_S^{n-1} [n]_{q_S} g_{1D} x_D^{n-1}.
\ee
Moreover $\Delta {x_D^n}_{\bb} = 0$, and it follows
\be
\Delta x_1^n {x_D^n}_{\bb} = q_S^{2(n-1)} [n]^2_{q_S} \mu g_{1D}
                        (x_1^{n-1} x_D^{n-1}),
\ee
thus
\be
<(t_1)^n (t_D)^n> = c_n^{-1} q_S^{n(n-1)} ([n]_{q_S}!)^2 (\mu g_{1D})^n
\ee
where $[n]_{q_S}! = [n]_{q_S} [n-1]_{q_S} ... [1]_{q_S}$.
Now $g_{D1} g_{1D} =1$, therefore
\be
((t_1)^n,(t_1)^n) = q_S^{n(n-1)}(\mu)^n c_n^{-1}([n]_{q_S}!)^2.
\ee
By another tedious, but straightforward calculation one finds
\be
c_n = \Delta^n {(x^2)^n}_{\bb} = \mu^{2n} q_S^{n(n-1) + n(D-2)/2}
    [n]_{q_S}![n+D/2-1]_{q_S}!([D/2-1]_{q_S}!)^{-1},
\label{c_n}
\ee
hence
\berr
((t_1)^n,(t_1)^n) &=& \mu^{-n} q_S^{-n(D-2)/2} [D/2-1]_{q_S}! [n]_{q_S}!
                      ([n+D/2-1]_{q_S}!)^{-1} \nn\\
             &=& (q_S^{(D-2)/2} + q_S^{-(D-2)/2})^{-n}
               \[\begin{array}{c} n+D/2-1 \\ D/2-1 \end{array}\]_{q_S}^{-1}, 
\err
where $\Big[ \begin{array}{c} n\\m \end{array} \Big]_{q} =
             \frac{[n]_{q}!}{[m]_{q}! [n-m]_{q} !}$. 
This is manifestly positive for $n < M_S-(D-2)/2 = k_S$.  $\quad \Box$


\begin{thebibliography}{9}
 
\bibitem{gomez} L. Alvarez--Gaum\'e, C. Gomez, G. Sierra, "Duality and Quantum 
 Groups" {\em Nucl. Phys.}  {\bf B330}, 347 (1990)
\bibitem{anderson} H.H. Andersen, P. Polo, W. Kexin, {\em Invent. math.}
  {\bf 104}, 1 (1991)
\bibitem{baez} J. Baez, ``An Introduction to Spin Foam Models of $BF$ Theory
  and Quantum Gravity'', gr-qc/9905087; L. Smolin, ``Quantum deformation of 
  quantum gravity'', gr-qc/9512020 
\bibitem{chang} Z. Chang, ``Quantum Anti--de Sitter space'', 
  hep-th/9904091
\bibitem{CH_P} V. Chari and A. Pressley, "A guide to quantum
  groups".   Cambridge University press, 1994
\bibitem{dobrev}  V. K. Dobrev, P. J. Moylan, {\em Phys. Lett.} {\bf 315B}, 
   292 (1993);
   V. K. Dobrev, R. Floreanini, {\em J. Phys. A}: Math. Gen. {\bf 27}, 
     4831 (1994); M. Flato, L.K. Hadjiivanov, I.T. Todorov,
   "Quantum Deformations of Singletons and of Free Zero-Mass Fields"
   {\em Foundations of Physics}, vol.{\bf 23} (4), 571-586 (1993)
\bibitem{drinfeld} V. Drinfeld, "Quantum Groups" {\em Proceedings
  of the International Congress of
  Mathematicians, Berkeley, 1986} A.M. Gleason (ed.), p. 798, AMS,
      Providence, RI
\bibitem{FRT}L.D.Faddeev, N.Yu.Reshetikhin, L.A.Takhtajan.
  "Quantization of Lie
  Groups and Lie Algebras"  {\em Algebra Anal.} {\bf 1} 178 (1989)
\bibitem{fiore} G. Fiore, "The $SO_q(N)$ - Symmetric Harmonic Oscillator on 
   the  Quantum Euclidean Space $R_q^N$ and its Hilbert Space Structure" 
   {\em Int. J. Mod. Phys} {\bf 26}, 4678 (1993)
\bibitem{ferrara} S. Ferrara, C. Fronsdal, ``Gauge fields and Singletons
  of $AdS_{2p+1}$'', hep-th/9806072
\bibitem{fronsdal} C. Fronsdal, {\em Rev. Mod. Phys.} {\bf 37},
  Nr. 1, 221 (1965);
  C. Fronsdal, {\em Phys. Rev. D} {\bf 10}, Nr. 2, 589 (1974);
  C. Fronsdal, R. Haugen, {\em Phys. Rev. D} {\bf 12}, Nr. 12, 3810 (1975);
  C. Fronsdal, {\em Phys. Rev. D} {\bf 20}, 848 (1979);
  J. Fang, C. Fronsdal, {\em Phys. Rev D} {\bf 22}, 1361 (1980)
\bibitem{weich} A. Hebecker and W. Weich, "Free Particle in $q$ - deformed 
           Configuration space". {\em Lett. Math. Phys.} 26, 245 (1992).
\bibitem{ho} P-M. Ho, S. Ramgoolam, R. Tatar, 
  ``Quantum Spacetimes and Finite $N$ Effects in 4D Super Yang--Mills 
  Theories'', hep-th/9907145
\bibitem{jantzen} J.C. Jantzen,  "Lectures on Quantum Groups". {\em
  Graduate Studies in Mathematics} Vol. 6, AMS 1996
\bibitem{ramgoolam} A. Jevicki, M. Mihailescu, S. Ramgoolam, ``Gravity
  from CFT on $S^N(X)$: Symmetries and Interactions'' hep-th/9907144;
  A. Jevicki, S. Ramgoolam, ``Non--commutative gravity from the 
  ADS/CFT correspondence'', hep-th/9902059
\bibitem{jimbo} M. Jimbo, "A q -- Difference Analogue of $U(g)$ and the
   Yang -- Baxter
 Equation" {\em Lett. Math. Phys} {\bf 10}, 63 (1985)
\bibitem{kachru} S. Kachru, E. Silverstein, ``4d Conformal Field Theories
  and Strings on Orbifolds'', hep-th/9802183
\bibitem{keller} G. Keller, "Fusion Rules of $U_q(sl(2,\compl))$, $q^m=1$".
      {\em Lett. Math. Phys.} {\bf 21}, 273 (1991)
\bibitem{kirill_resh} A.N. Kirillov, N. Reshetikhin,  "q- Weyl group
  and a Multiplicative Formula for Universal R- Matrices" {\em Comm.
  Math. Phys.} {\bf 134}, 421 (1990)
\bibitem{tolstoi} S.M. Khoroshkin, V.N. Tolstoy, "Universal R- Matrix
  for    Quantized (Super)Algebras" {\em Comm. Math. Phys} {\bf 141}, 599
   (1991)
\bibitem{lev_soib} S.Z. Levendorskii, Y.S. Soibelman, "Some
  applications of the quantum Weyl group" {\em Journ. Geom. Phys} {\bf 7}
  (2), 241 (1990)
\bibitem{lusztig} G. Lusztig, "Quantum deformations of certain simple
  modules ...", {\em Adv. in Math.} {\bf 70}, 237 (1988);  G. Lusztig, "On
  quantum groups" {\em J. Algebra} {\bf 131}, 466(1990);
  G. Lusztig,  "Quantum Groups at roots of 1"
  {\em Geom. Ded.} {\bf 35}, 89 (1990)
\bibitem{lusztig_book}  G. Lusztig,  "Introduction to Quantum
  Groups". {\em Progress in Mathematics} Vol. 110, Birkhaeuser 1993
\bibitem{mack_schom} G. Mack, V. Schomerus,  "Quasi Hopf quantum
  symmetry in quantum theory" {\em Nucl. Phys.} {\bf B 370}, 185 (1992)
\bibitem{madore} J. Madore, ``The fuzzy sphere'' {\em Class. and Quantum
  Gravity} {\bf 9}, 69 (1992); H. Grosse, C. Klim\v{c}ik, P. Pre\v{s}najder, 
  ``Towards Finite Quantum Field Theory
  in Non--Commutative Geometry'' {\em Int. Journ. Mod. Phys.} {\bf A 35}, 231
  (1996)
\bibitem{maldacena} J. Maldacena, ``The large $N$ limit of superconformal
  field theories and supergravity'', {\em Adv. Theor. Math. Phys.} 
  {\bf 2}, 231 (1998); hep-th/9711200
\bibitem{strominger} J. Maldacena, A. Strominger, ``AdS3 Black Holes and
  a Stringy Exclusion Principle'' {\em JHEP} {\bf 9812}, 5 (1998); 
  hep-th/9804085
\bibitem{oeckl} R. Oeckl, ``Braided Quantum Field Theory''; hep-th/9906225
\bibitem{paradowski} J. Paradowski, "Filtrations of modules over the Quantum 
   Algebra",  {\em Proc. Symp. Pure Math.} {\bf 56}, 93 (1994) 
\bibitem{resh_1} N. Reshetikhin, "Quantized universal enveloping algebras, 
  the Yang--Baxter equation and invariants of links I,II" 
  LOMI preprint E-4-87, E-17-87. 
\bibitem{rosso} M. Rosso, "Finite Dimensional Representations of the Quantum 
  Analog of the Enveloping Algebra of a Complex Simple Lie Algebra" 
   {\em Comm. Math. Phys.} {\bf 117}, 581 (1988)
\bibitem{schmuedgen} K. Schm\"udgen, ``Operator \reps of 
  ${\cal U}_q(sl_2(\R))$'',  {\em Lett. Math. Phys}. {\bf 37}, 211 (1996)
\bibitem{soib} Y.S. Soibelman, "Quantum Weyl group and some of its
  applications"
  {\em Rend. Circ. Mat. Palermo Suppl} {\bf 26} (2), 233 (1991)
\bibitem{integral} H. Steinacker, "Integration on quantum Euclidean space
  and sphere"
 {\em J. Math. Phys.} {\bf 37} (Nr.9), 7438 (1996)
\bibitem{thesis} H. Steinacker, "Quantum Groups, Roots of Unity and Particles
  on quantized Anti--de Sitter Space". Ph.D. Thesis, Berkeley, May 1997; 
  {\em hep-th/9705211} 
\bibitem{brst}  H. Steinacker, ``Unitary Representations and BRST Structure 
  of the Quantum Anti--de Sitter Group at Roots of Unity`` 
  Proceedings to WigSym5, Vienna, Austria, 25-29 August, 1997; q-alg/9710016
\bibitem{nc_reps} H. Steinacker, ``Unitary Representations of Noncompact
 Quantum Groups at Roots of Unity'', math.QA/9907021
\bibitem{ads_paper} H. Steinacker, "Finite--dimensional Unitary 
  Representations of quantum Anti--de Sitter Groups at Roots of Unity",
  {\em Comm. Math. Phys.} {\bf 192}, 687 (1998)
\bibitem{WZ} J. Wess, B. Zumino, "Covariant Differential Calculus
  on the Quantum Hyperplane"  {\em Nucl Phys} {\bf 18B} (Proc. Suppl.),
  302 (1990); U. Carow--Watamura, M. Schlieker, S. Watamura, 
  {\em Z. Phys. C  Part. Fields} {\bf 49}, 439 (1991)
\bibitem{duff} B. Zumino, {\em Nucl. Phys.} {\bf B 127}, 189 (1977);
   S. Deser, B. Zumino, {\em Phys. Rev. Lett} {\bf 38}, 1433 (1977);
   M.J. Duff, B.E.W. Nilsson, C.N. Pope, ``Kaluza--Klein Supergravity'',
  {\em Phys. Rep.} {\bf 130}, Nr. 1 \& 2 (1986)
\bibitem{zumino_priv} B. Zumino, private communication

\end{thebibliography}
\end{document}